\documentclass[aps,pra,twocolumn,amsmath,amssymb,nofootinbib,showpacs,superscriptaddress]{revtex4-1}
\usepackage[english]{babel}
\usepackage{latexsym}
\usepackage{graphics}
\usepackage{graphicx}
\usepackage{epsfig}
\usepackage{color}
\usepackage{bm}
\usepackage{amsmath}
\usepackage{amssymb}
\usepackage{amsthm}
\usepackage{dcolumn}
\usepackage{bm}
\usepackage{float}
\usepackage{hyperref}
\usepackage{color}
\usepackage{epstopdf}
\usepackage{cleveref, braket, comment,bbold}
\usepackage[svgnames]{xcolor}
\hypersetup{hidelinks,colorlinks=true,allcolors=DarkBlue}
\usepackage[normalem]{ulem}

\newcommand{\bin}{\operatorname{bin}}
\newcommand{\one}{\ensuremath{\ket{\underline{1}}}}
\newcommand{\zero}{\ensuremath{\ket{\underline{0}}}}

\theoremstyle{remark}

\begin{document}

\preprint{APS/123-QED}

\title{Efficient realization of quantum algorithms with qudits}

\author{Anastasiia S. Nikolaeva}
\affiliation{Russian Quantum Center, Skolkovo, Moscow 121205, Russia}
\affiliation{National University of Science and Technology ``MISIS”,  Moscow 119049, Russia}

\author{Evgeniy O. Kiktenko}
\affiliation{Russian Quantum Center, Skolkovo, Moscow 121205, Russia}
\affiliation{National University of Science and Technology ``MISIS”,  Moscow 119049, Russia}

\author{Aleksey K. Fedorov}
\affiliation{Russian Quantum Center, Skolkovo, Moscow 121205, Russia}
\affiliation{National University of Science and Technology ``MISIS”,  Moscow 119049, Russia}

\date{\today}
\begin{abstract}
The development of a universal fault-tolerant quantum computer that can solve efficiently various difficult computational problems is an outstanding challenge for science and technology.
In this work, we propose a technique for an efficient implementation of quantum algorithms with multilevel quantum systems (qudits). 
Our method uses a transpilation of a circuit in the standard qubit form, which depends on the characteristics of a qudit-based processor, such as the number of available qudits and the number of accessible levels. 
This approach provides a qubit-to-qudit mapping and comparison to a standard realization of quantum algorithms highlighting potential advantages of qudits.
We provide an explicit scheme of transpiling qubit circuits into sequences of single-qudit and two-qudit gates taken from a particular universal set.
We then illustrate our method by considering an example of an efficient implementation of a $6$-qubit quantum algorithm with qudits. 
In this particular example, we demonstrate how using qudits allows a decreasing amount of two-body interactions in the qubit circuit implementation.
We expect that our findings are of relevance for ongoing experiments with noisy intermediate-scale quantum devices that operate with information carriers allowing qudit encodings, 
such as trapped ions and neutral atoms, as well as optical and solid-state systems. 
\end{abstract}

\maketitle

\section{Introduction}

Progress in engineering coherent quantum many-body systems with a significant degree of control makes it realistic to study properties of exotic quantum phases~\cite{Lukin2017,Lukin2021,Monroe2017,Browaeys2018,Martinis2019,Blatt2018-2} 
and to prototype quantum algorithms~\cite{Martinis2016-2,Gambetta2017-2,Blatt2018,Gambetta2018}. 
One of the key issue in the future scaling of such systems is preserving their coherent properties when the system size is increased.  
Existing prototypes of quantum computing devices are based on various physical platforms, such as superconducting circuits~\cite{Martinis2019}, 
semiconductor quantum dots~\cite{Vandersypen2022,Morello2022,Tarucha2022}, 
trapped ions~\cite{Monroe2017,Blatt2018,Blatt2018-2}, 
neutral atoms~\cite{Lukin2017,Lukin2021,Browaeys2018}, 
photons~\cite{Pan2020,Madsen2022}, etc. 
The use of such objects as two-level systems (qubits) in many cases is an idealization since underlying physical systems are essentially multilevel. 
The idea of using additional levels of quantum objects for realizing quantum algorithms is at the heart of {\it qudit-based} quantum information processing.
This approach has been widely studied last decades~\cite{kiktenko2023realization,Zeilinger2018}, 
both theoretically and experimentally~\cite{Farhi1998,Kessel1999,Kessel2000,Kessel2002,Muthukrishnan2000,Nielsen2002,Berry2002,Klimov2003,Bagan2003,Vlasov2003,Clark2004,Leary2006,Ralph2007,White2008, Zobov2008, Ionicioiu2009,Ivanov2012,Li2013,Kiktenko2015,Kiktenko2015-2, Song2016,Frydryszak2017,Bocharov2017,Gokhale2019,Pan2019,Low2020,Jin2021,Martinis2009,White2009,Wallraff2012,Mischuck2012,Gustavsson2015,Martinis2014,Ustinov2015, Morandotti2017,Balestro2017,Low2020,Sawant2020,Pavlidis2021,Rambow2021, Zobov2022, Vashukevich2022, Goss2022, Gonzalez2022}.
The most recent result is the pioneering realization of universal multi-qudit processors with trapped ions~\cite{Ringbauer2021, Aksenov2022, decoupling2023fian}, superconducting~\cite{Hill2021, Goss2022} and optical systems~\cite{Chi2022}.

Although manipulating with additional levels faces additional
challenges, recent experiments show dramatic progress in increasing the fidelities of qudit operations and making them comparable with the ones for qubits. 
In particular, high-fidelity qutrit $\sf CZ$ and $\sf CZ^\dagger$ gates, with estimated process fidelities of 97.3(1)\% and 95.2(3)\%, respectively, have been recently demonstrated in Ref.~\cite{Goss2022}. 
Also with superconducting systems, fidelity of 97.7\% for two-qutrit $\sf CPHASE$ gate have been achieved \cite{Hill2021}. 
Two qudit $\sf MVCXd$ gate on two photonic ququarts has been implemented with fidelity 95.2\% in \cite{Chi2022}. 
For the trapped ion platform, on which a qudit processor with 8-level qudits was developed, two-qutrit gate fidelity 97.5(2)\% has been achieved~\cite{Ringbauer2021}.
Remarkably, 8-level qudits are controlled by a single laser acousto-optic modulator (AOM) as reported in Ref.~\cite{Ringbauer2021} 

Quantum algorithms within the digital quantum computing model can be presented as qubit-based circuits, so there are several approaches for processing them using qudits.
First of all, qudits can be decomposed of a set of qubits~\cite{Kessel1999,Kessel2000,Kessel2002,Kiktenko2015,Kiktenko2015-2, popov2016information}. 
This approach may decrease the cost of realizing quantum algorithms by replacing some two-qubit operations requiring interaction between distinct physical objects by single-qudit ones, which do not require an interaction between distinct physical objects.
However, this method is not universal in the sense that the total number of operations strongly depends on the {\it mapping}, {\it i.e.} the way how qubits are encoded in qudits. 
As we demonstrate below, specific mappings applied to specific qubit circuits may even lead to a substantial increase in the number of operations in comparison with the standard qubit-based approach.
Second, higher qudit levels can be used for substituting ancilla qubits~\cite{Ralph2007,Gokhale2019,Kiktenko2020,Nikolaeva2022}. 
This is especially important for decomposing multiqubit  gates, such as the generalized Toffoli gate.
In particular, an additional (third) energy level of a transmon qubit has been used in the experimental realization of Toffoli gate~\cite{Wallraff2012} (see also Refs.~\cite{Galda2021,Gu2021}), 
which is a key primitive of many quantum algorithms, such as Shor's and Grovers's algorithms. 
While existing quantum computing schemes that are based on qubits platform benefit from several approaches for the realization of quantum algorithms, which require compilers, transpilers, and optimizers, 
qudit-based quantum computing remains described mostly at the level of logic operations~\cite{Chong2017}.

In the present work, we propose a technique for an efficient realization of qubit-based quantum algorithms,
which employs the combination of two aforementioned approaches for the use of additional levels of qudits. 
The crucial element of our method is a transpilation of a qubit circuit, which depends on the parameters of an accessible qudit-based processor (e.g., number of levels and fidelity of operations). 
As a result, one obtains qubit-to-qudit mapping and comparison to the standard qudit realization.
A qudit circuit can be executed via quantum processors or classical emulators, and corresponding outcomes can be further post-processed in order to be interpreted as results of an algorithm.
Clearly, due to exponential complexity, classical emulation is possible only in the case of low-width  or low-depth circuits.
We develop an explicit scheme of transpiling qubit circuits into sequences of single-qudit and two-qudit gates taken from a particular universal set, which can be different for quantum processors based on various physical platforms.
We provide an illustrative example of the qudit-based transpilation for a six-qubit quantum circuit, where we demonstrate the main features of our approach.
We also discuss types of quantum algorithms, where the developed approach can show the greatest improvement compared with a straightforward qubit-based implementation.

The paper is organized as follows.
In Sec.~\ref{sec:qubit}, we revise the basic principles of quantum computing with qubits.
In Sec.~\ref{sec:qudits}, we discuss the general approach for implementing qubit circuit on qudit-based processors.
In Sec.~\ref{sec:transpiler-description}, we provide a concrete realization of a qudit-based transpiler.
In Sec.~\ref{sec:example}, we present an example of applying the developed approach for realizing 6-qubit circuits with four 4-level qudits.
In Sec.~\ref{sec:disc}, we discuss the scalability of the developed approach and its most promising use cases.
Finally, we conclude in Sec.~\ref{sec:concl}.

\section{Qubit-based approach} \label{sec:qubit}

The essence of qubit-based quantum computation is applying a unitary operator $U^{\rm qb}_{\rm circ}$ to a set of $n$ two-level particles (qubits), initialized in the fixed state $\ket{0}^{\otimes n}$, 
and the measuring the resulting state in the computational basis to obtain a sample from the following distribution:
\begin{equation}\label{eq:distr}
	p^{\rm {qb}}(x)=\left|\braket{x|U_{\textrm{circ}}^{\textrm{qb}}|0}^{\otimes n}\right|^2.
\end{equation}
Here we denote computational basis states of qubits as $\ket{0}$ and $\ket{1}$, 
$\ket{x}\equiv \ket{x_0}\otimes\ldots\otimes\ket{x_{n-1}}$, and $x=(x_0,\ldots,x_{n-1})\in\{0,1\}^n$.
Commonly, the same circuit is executed several times, which results in a sequence of independent and identically distributed (i.i.d.) random $n$-bit strings $(x^{(1)},\ldots,x^{(N)})$, 
where $N$ is the number of samples, and each sample $x^{(i)}$ is obtained from distribution~\eqref{eq:distr}.

The operator  $U^{\rm qb}_{\rm circ}$ is originally represented in the form of a sequence of some standard unitary operators (gates) $U_i^{\rm qb}$ constituting hardware-agnostic (idealized)  circuit ${\sf circ}^{\rm qb}$.
For applying $U^{\rm qb}_{\rm circ}$ to real physical objects, 
an additional \emph{transpilation} step of decomposing $U^{\rm qb}_{\rm circ}$ 
to \emph{native} (usually, single-qubit and two-qubit) operations is required~\cite{Chong2017,Monroe2021,Earnest2021}.
One of DiVincenzo's criteria~\cite{DiVincenzo2000} to quantum processors is the requirement to realize a universal set of gates that allows obtaining an efficient approximation of an arbitrary unitary operation up to a predefined accuracy.
Although multiqubit processors based on various physical principles have been demonstrated, the problem limited quality of quantum operations restricts the computational capabilities of such systems 
\cite{Fedorov2022}.
A particular issue is the realization of high-quality two-qubit quantum operations that require interactions between quantum information carriers.
Another important factor that has to be taken into account, is the restricted coupling map of information carriers, which represents the opportunity to implement two-body interactions.
This issue can be overcome by adding additions $\sf SWAP$ operations. 
However, this problem is beyond the scope of our work, and further we suppose that the quantum processor has an all-to-all coupling map.

\section{Quantum computing with qudits} \label{sec:qudits}

The idea of using of qudits, i.e. $d$-level quantum systems with $d>2$, have been widely considered in the context of quantum information processing~\cite{Farhi1998,Kessel1999,Kessel2000,Kessel2002,Muthukrishnan2000,Nielsen2002,Berry2002,Klimov2003,Bagan2003,Vlasov2003,Clark2004,Leary2006,Ralph2007,White2008,Ionicioiu2009,Ivanov2012,Li2013,Kiktenko2015,Kiktenko2015-2, Song2016,Frydryszak2017,Bocharov2017,Gokhale2019,Pan2019,Low2020,Jin2021,Martinis2009,White2009,Wallraff2012,Mischuck2012,Gustavsson2015,Martinis2014,Ustinov2015, Morandotti2017,Balestro2017,Low2020,Sawant2020,Pavlidis2021,Rambow2021}.
Clearly, an $m$-qudit system can be used in order to obtain the same result as in the case of qubit-based computing 
-- obtaining the number of samples coming from the distribution determined by an $n$-qubit circuit, but potentially with fewer resources, e.g. smaller number of information carriers and/or operations.
The dimension of qudits and their number has to be compatible with the given $n$-qubit circuit. 
In what follows, we assume that $d^m \geq 2^n$.

\begin{figure}
\center{\includegraphics[width=\linewidth]{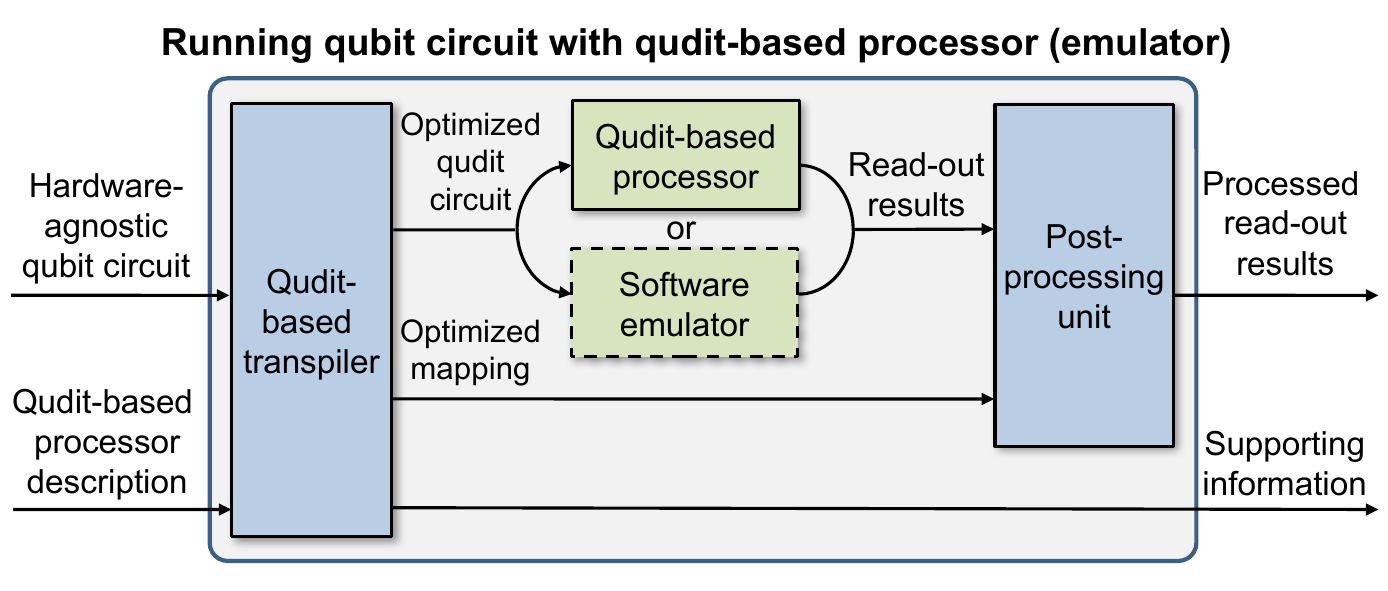}}
\vskip -4mm
\caption{The main stages of implementing a qubit circuit with a qudit-based processor or emulator.}
\label{fig:scheme}
\end{figure}

A specific question that we are interested in is a transpilation of a circuit given in the qubit form depending on the parameters of a qudit-based processor.
A scheme of our approach is presented in Fig.~\ref{fig:scheme}.
It consists of the three following stages: (i) qubit-to-qudit circuit transpilation, (ii) circuit execution, and (iii) classical post-processing of the measurement results.
We note that stages (i) and (iii) are performed with a classical computer, while stage (ii) is realized with an accessible qudit-based processor or its classical software emulator.

The input for our scheme is a hardware-agnostic qubit circuit ${\sf circ}^{\rm qb}$, necessary runs number $N$, and the general information about the accessible qudit-based processor, 
specifically, the number of qudits $m$, their dimension $d$, and the set of native gates 
(usually, it consists of single-qudit gates and a set of two-qudit gates within a certain connectivity graph indicating the possibility of direct realization of a two-qubit/two-qudit operation).
The output of the transpilation step is an `optimized' qudit circuit ${\sf circ}^{\rm qd}_{\rm opt}$, which is a sequence of native qudit gates, and an `optimized qubit-to-qudit mapping' that is an injective function
\begin{equation}
    \phi_{\rm opt} : \{0,1\}^n \rightarrow \{0,1,\ldots,d-1\}^m,
\end{equation}
assigning a qudit's computational basis state to each of the qubits' ones.
The general idea is that running of ${\sf circ}^{\rm qd}_{\rm opt}$ and processing the output measurement outcomes according to $\phi_{\rm opt}$ provides bitstrings equivalent to ones obtained 
after running ${\sf circ}^{\rm qb}$ on a standard qubit-based quantum processor (we formulate the rigorous consistency condition below).
The term optimized appears here because various qubit-to-qudit mappings, which are assignments between qubits' and qudits' levels, result in different qudit-based circuits that are equivalent to the input qubit-based circuit under a particular mapping.
In this way, the goal of the qudit-based transpiler consists not only in transforming qubit gates of ${\sf circ}^{\rm qb}$ to native qudit ones but also in finding a favorable mapping such that the realization of the resulting qudit-based circuit is beneficial over the straightforward realization of ${\sf circ}^{\rm qb}$ on a qubit-based processor.
We note that the optimized mapping depends both on the input circuit (it can be different for different circuits) and the architecture of the accessible qudit-based processor.

The desirable characteristics of the mapping can be defined in different ways.
Below, we consider a particular implementation of a qudit-based transpiler, where we use the number of two-qudit interactions as the main figure of merit for quantifying the performance of the transpilation (see Sec.~\ref{sec:transpiler-description}).
The reason for this is that usually, two-body gates are the main source of errors during the process of executing quantum circuits.
Nevertheless, alternative metrics, such as circuit depth or resulting fidelity estimation, can be used.

To achieve the goal of reducing the number of two-body gates in going from qubits to qudits, two main techniques, as well as their combination, can be implemented.
The first technique~\cite{Kessel1999,Kessel2000,Kessel2002,Kiktenko2015,Kiktenko2015-2}, employs a qudit's $d$-dimensional space for embedding several qubits (the technique works for $d\geq 2^{m'}$ with $m'>1$).
Its main advantage is the possibility to reduce the number of employed physical information carriers (e.g., particles, such as atoms or ions).
However, as we show in Sec.~\ref{sec:transpiler-description}, this method is not universal in the sense that the total number of operations strongly depends on the mapping, i.e. the way how qubits are embedded in qudits. 
It appears that the cost of the realization of two-qubit operations between two qubits inside one qudit may be close to a couple of single-qudit operations, since it does not require any interaction between distinct physical particles.
In contrast, in the realization of a two-qubit operation between qubits belonging to different qudits, additional entangling operations are required to presume the state of other qubits inside these qudits but untouched by the two-qubit gate.

The second technique is to use `upper' qudit levels ($\ket{\rm a}$, $a\geq 2$) for substituting ancillary qubits within standard multiqubit gates decompositions~\cite{Ralph2007,White2009,Gokhale2019,Kiktenko2020,Nikolaeva2022}.
This approach allows decreasing both the number of required two-body interactions (entangling gates) and the number of employed quantum information carriers by removing the necessity of ancillary qubits and is useful in the case of quantum circuits containing multiqubit gates.
We would like to note that these two approaches can be combined in the case of $d>2^{t}$ for some $t\geq 2$: 
The first $2^t$ levels of a qudit can be used for embedding $\lfloor\log_2 d\rfloor$ qubits, while the remaining ones can be used for subsisting ancillas.

There are two main aspects regarding the qudit-based transpilation.
The first is related to the possibility of realizing qudit gates. 
As for qubits, a universal set of gates can be composed of arbitrary single-qudit gates, supplemented with a two-qudit entangling gate of a particular type.
One of the approaches for making this two-qudit gate is to employ the original two-qubit gate (used within the qubit-based architecture), yet considered in the full qudit state space.
We note that this approach has been successfully demonstrated in experiments with trapped ions, and it has been shown that the resulting gate fidelities are comparable with the ones for corresponding qubit-based architectures~\cite{Ringbauer2021, Aksenov2022}.

The second aspect is related to finding an appropriate qubit-to-qudit mapping.
In the case of small- and intermediate-scale circuits, one may use an exhaustive search through all possible mappings. 
However, this approach requires significant classical computational resources for large-scale circuits.
In this case, one is sufficed to find a mapping that is not the best possible one, but still gives the lower number of two-body gates compared to the standard qubit implementation (or gives a higher fidelity).
If the number of available qudits $m$ is not less than the number of qubits in the input circuit $n$, then it can be assured that the number of two-qudit gates in the resulting qudit circuit does not exceed the number of two-qubit gates in the input circuit.
This follows from the fact that there is a trivial mapping, where each qubit is embedded in its own qudit. 
In the qutrit case ($d=3$) and $m\geq n$, there is no problem with searching for the appropriate mapping:
One can employ $n$ qutrits, each used as a qubit plus the ancillary state.
For more complex embeddings of qubits in qudits we describe several approaches of the optimized mapping finding algorithms in Sec.~\ref{sec:mappingfinder}.
The comparison between the number of two-body gates for the best-found mapping with the number of two-body gates for the straightforward qubit-based implementation can serve as a benchmark for the efficiency of the qudit-based transpilation process and could be placed in the supporting information.

Let us back to the description of the main stages of running the qubit-based circuit with the qudit-based processor shown in Fig.~\ref{fig:scheme}.
At stage (ii), the qudit circuit ${\sf circ}^{\rm qd}_{\rm opt}$ is the input for the qudit-based processor (or emulator) that applies the gates from  ${\sf circ}^{\rm qd}_{\rm opt}$  to the qudit register initialized in the state $\ket{0}^{\otimes m}$,
where we use $\{\ket{l}\}_{l=0}^{d-1}$ to denote computational basis states of each qudit.
The resulting qudit state is measured in the computational basis, and a sample from the following distribution is obtained:
\begin{equation}\label{eq:qudit-distr}
	p^{\rm {qd}}(y)=\left|\braket{y|U_{\textrm{circ}}^{\textrm{qd}}|0}^{\otimes m}\right|^2,
\end{equation}
where $U_{\textrm{circ}}^{\textrm{qd}}$ is the resulting qudit unitary operator, $y=(y_0,\ldots,y_{m-1})$, $y_i\in\{0,\ldots,d-1\}$, and $\ket{y}\equiv \ket{y_0}\otimes\ldots\otimes\ket{y_{m-1}}$.
The circuit is run $N$ times, that yields a $N$-length sequence $(y^{(1)},\ldots,y^{(N)})$, where each $y^{(i)}$ is the string of $m$ numbers from $\{0,\ldots,d-1\}$.

The final post-processing stage takes the read-out results $(y^{(1)},\ldots,y^{(N)})$ and a mapping $\phi$ in order to obtain equivalent qubit outcomes $(\phi^{-1}(y^{(1)}),\ldots,\phi^{-1}(y^{(N)}))$ as output, 
where $\phi^{-1}$ outputs $n$-length bit strings out of $y^{(i)}$.
The general condition for the scheme's correctness is as follows:
\begin{equation} \label{eq:consistency}
	p^{\rm qd}(y) = p^{\rm qb}(\phi^{-1}(y)),\quad y\in {\rm image}(\phi),
\end{equation}
where ${\rm image}(\phi)$ is the set of all possible outputs of the mapping $\phi$.
The consistency condition~\eqref{eq:consistency} guarantees that  only $y\in {\rm image}(\phi)$ can appear as measurements results of the qudit circuit, 
and the obtained bit strings $(\phi^{-1}(y^{(1)}),\ldots,\phi^{-1}(y^{(N)}))$ are indistinguishable from ones that can be obtained with a qubit-based processor.
The set of bitstrings $(\phi^{-1}(y^{(1)}),\ldots,\phi^{-1}(y^{(N)}))$ together with the supporting information is the final output of our approach.
One can see that from the viewpoint of classical processing, the most challenging is the qudit-based transpilation stage.
We discuss it in detail below.

\section{Qudit-based transpilation}\label{sec:transpiler-description}
Here we describe the concrete realization of the qudit-based transpiler designed for a specific model of a qudit-based processor.
We assume that the available processor consists of $m$ $d$-dimensional qudits, labeled as ${\sf Q}1,\ldots,{\sf Q}m$.
As a set of native qudit gates, we consider single-qudit operations
\begin{equation}   
	\begin{split}
		&R_{{\sf Q}j}^{\alpha,\beta}(\varphi,\theta) = \exp(-\imath \theta \sigma^{\alpha,\beta}_{\varphi}/2),\\
		&{\rm Ph}_{{\sf Q}j}^{\alpha}(\theta) = \text{diag}(1,\dots,1,e^{\imath \theta}, 1,\dots,1),
	\end{split}\label{eq:sqd-gates}
\end{equation}
where $e^{\imath \theta}$ is located in $\alpha$th position,
and a two-qudit operation
\begin{equation}\label{eq:qudit-cz}
	{\sf CZ}^{\alpha, \beta}_{{\sf Q}j_1, {\sf Q}j_2} = \mathbb{1}_{{\sf Q}j_1}\otimes \mathbb{1}_{{\sf Q}j_2}-2\ket{\alpha\beta}_{{\sf Q}j_1, {\sf Q}j_2}\bra{\alpha\beta},
\end{equation}
{which applies a fixed phase factor $-1$ to the pair of levels given by $\alpha$ and $\beta$}.
Here we use the following notations:
\begin{equation}
	\begin{aligned}
	&\sigma_\varphi^{\alpha,\beta}=\sigma_{x}^{\alpha,\beta}\cos\varphi  + \sigma_{y}^{\alpha,\beta}\sin\varphi,\\
		&\sigma^{\alpha,\beta}_{\xi} = 
		\bra{0}\sigma_{\xi}\ket{0}\!\ket{\alpha}\!\bra{\alpha}+
		\bra{0}\sigma_{\xi}\ket{1}\!\ket{\alpha}\!\bra{\beta}+\\
		&\bra{1}\sigma_{\xi}\ket{0}\!\ket{\beta}\!\bra{\alpha}+
		\bra{1}\sigma_{\xi}\ket{1}\!\ket{\beta}\!\bra{\beta}, ~\xi \in\{x, y, z\},
	\end{aligned}
\end{equation}
$\sigma_x$, $\sigma_y$, $\sigma_z$ are standard single-qubit Pauli matrices, $\alpha,\beta \in \{0,\ldots, d-1\}$ denote levels in qudits' space, $\varphi,\theta$ are real-valued arbitrary angles, and $\mathbb{1}$ stands for the identity matrix.
In what follows, we use subindices over unitary operators to specify quantum information carriers or carriers (qubits or qudits) on which this operator acts.
We assume that two-qudit gates can be implemented for every pair of qudits within the all-to-all coupling map.
{
Note that ${\sf CZ}^{1,1}_{{\sf Q}j_1, {\sf Q}j_2}$ realizes a standard qubit controlled-phase gate acting in the four-dimensional subspace spanned by the first two levels of ${\sf Q}j_1$ and ${\sf Q}j_2$, and acts as identity in the remaining subspace.
Moreover, ${\sf CZ}^{\alpha,\beta}_{{\sf Q}j_1, {\sf Q}j_2}$ with arbitrary $\alpha$ and $\beta$ can be realized by surrounding a single instance of ${\sf CZ}^{1,1}_{{\sf Q}j_1, {\sf Q}j_2}$ with single-qudit operations.
}

We note that to realize the considered single-qudit gates in Eq.~\eqref{eq:sqd-gates}, it is enough to have a connected (but not fully connected) coupling graph of allowed transitions between levels, as shown in Ref.~\cite{Mato2022}.
Knowing the exact coupling map between levels, single-qudit operations can be easily reformulated in terms of accessible transitions. 
This is the case for superconducting \cite{Blok2021, kazmina2023demonstration}, ion-based \cite{Ringbauer2021, Low2020, Aksenov2022}, and neutral-atom-based \cite{Gonzalez2022} qudits. 
Moreover, in real existing experimental setups, transitions within a given coupling graph are usually addressed with a single laser. 
For example, in Ref.~\cite{Ringbauer2021}, 10 allowed transitions inside 8-level qudit realized by $^{40}{\rm Ca}^+$ ions are accessed by a single narrowband laser at 729nm with AOM.
The employed two-qudit gate (\ref{eq:qudit-cz}) can be realized via Rydberg blockade neutral atom-based \cite{Gonzalez2022} qudits, and via common quantized motion mode in ion-based platform \cite{Cirac1995}.

\begin{figure}
	\centering
	\includegraphics[width=0.8\linewidth]{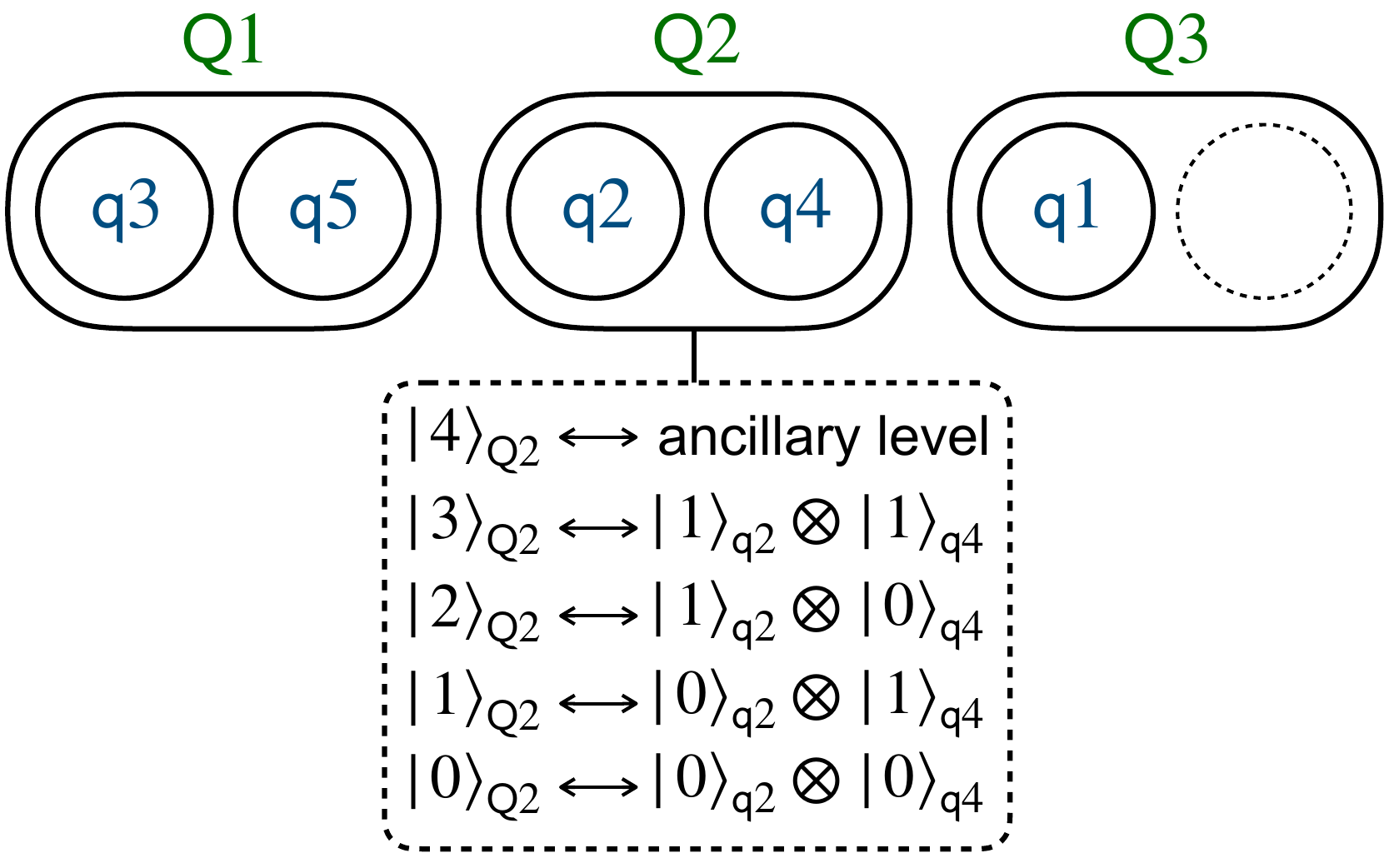}
	\caption{Example of a mapping in the form, given by Eq.~\eqref{eq:mapping} and~\eqref{eq:binarymapping}, between $n=5$ qubits and $m=3$ qudits of dimension $d=5$. 
	Within the presented mapping, 
	${\sf position{\_}in{\_}qudit[{\sf q}4]}=2$,
	${\sf qudit{\_}index}[{\sf q}4]={\sf Q}2$, and
	${\sf qubit{\_}index}[{\sf Q}2, 2]={\sf q}4$.}
	\label{fig:qubit-qudit-mapping}
\end{figure}

The input for the designed transpiler $n$-qubit hardware-agnostic qubit circuit ${\sf circ}^{\rm qb}$, acting of qubits denoted by ${\sf q}1,\ldots,{\sf q}n$, is assumed to consist of single-qubit gates
\begin{equation}\label{eq:qubit_rotation}
	r_{{\sf q}i}(\varphi,\theta) = \exp(-\imath \theta \sigma_{\varphi}/2),
\end{equation}
where $\sigma_\varphi=\sigma_x\cos\varphi+\sigma_y\sin\varphi$, and a $\kappa$-qubit gates
\begin{multline}\label{eq:qubit-cz}
	{\sf CZ}_{{\sf q}i_1,\ldots, {\sf q}i_\kappa}\!= \!\mathbb{1}_{{\sf q}i_1}\!\otimes\dots\otimes\! \mathbb{1}_{{\sf q}i_\kappa}\!-\\
	-\!2\ket{1\dots1}_{{\sf q}i_1,\ldots, {\sf q}i_\kappa}\!\bra{1\dots 1}
\end{multline}
with $\kappa\in\{2,3,\ldots,n\}$.
One can see that multi-body operations~\eqref{eq:qubit-cz} and~\eqref{eq:qudit-cz} correspond to acquiring a phase factor of $-1$ on a particular multi-body state.
We note that a multi-qubit operation~\eqref{eq:qubit-cz} can be transformed into a generalized Toffoli gate by applying single-qubit gates.
We also note that both the considered qudit-based and qubit-based sets of gates are universal.

Without loss of generality, we assume that ${\sf circ}^{\rm qb}$ terminates with read-out measurements in a computational basis acting on each of $n$ qubits.
The initial state of the qubit register is assumed to be $\ket{0}^{\otimes n}$.
We note, however, that the developing technique of transforming qubit gates into qudit gates is independent of the chosen initial state, and can be applied in the same way within other types of initialization.

For possible mappings between qubits' and qudits' spaces, we restrict ourselves with embeddings of the whole computational space of one or several qubits into a space of a particular qudit.
Specifically, each possible mapping $\phi$ can be defined in the following form:
\begin{equation}\label{eq:mapping}
    \phi = (\phi_{{\sf Q}1},\ldots, \phi_{{\sf Q}{m}}),\quad
    \phi_{{\sf Q}j} = ({\sf q}i_{j,1},\ldots,{\sf q}i_{j,\#j}),
\end{equation}
meaning that qudit ${\sf Q}j$ contains $\#j$ qubits ${\sf q}i_{j,1},\ldots,{\sf q}i_{j,\#j}$ (given that the condition $2^{\#j}\leq d$ is fulfilled).
The assignment of computational basis states of qubits to computational basis states of qudits is governed by the corresponding binary representation:
\begin{equation}\label{eq:binarymapping}
    \ket{ {\rm int}(x_1,\ldots,x_{\#j}) }_{{\sf Q}j} \leftrightarrow 
    \ket{ x_1 }_{{\sf q}i_{j,1}}\otimes \ldots \otimes 
    \ket{ x_{\#j} }_{{\sf q}i_{j,\#j}},
\end{equation}
where $x_1,\ldots, x_{\#j}\in\{0,1\}$ are bit values and ${\rm int}(x_1,\ldots,x_{\#j})\in\{0,1,\ldots,2^{\#j}-1\}$ outputs an integer number from its binary representation $x_1,\ldots, x_{\#j}$.
We note that within the considered mappings, the `address' of each qubit is defined by an index of qudit ${\sf Q}j\in\{{\sf Q}1,\ldots,{\sf Q}m\}$ and a `position' ${\sf pos}\in\{1,\ldots,\#j\}$ of the qubit inside the qudit.
An example of a possible mapping between $n=5$ qubits and $m=3$  qudits of the dimension $d=5$ is shown in Fig.~\ref{fig:qubit-qudit-mapping}.

To simplify operations within the considered special case of mapping $\phi$, it is convenient to introduce the following functions:
\begin{equation}
	\begin{split}
		&{\sf position{\_}in{\_}qudit}[{\sf q}i]\mapsto {\sf pos},\\
		&{\sf qudit{\_}index}[{\sf q}i]\mapsto{\sf Q}j,\\
		&{\sf qubit{\_}index}[{\sf Q}j, {\sf pos}]\mapsto{\sf q}i,
	\end{split}
\end{equation}
where the first two functions provide the address of a given qubit with the qudits' space, and the third function returns an index of a qubit given its address.
We also introduce the following function:
\begin{equation}
    \begin{split}
        & {\sf indices{\_}of{\_}qubits}[{\sf Q}j] \rightarrow
        \{{\sf q}i_{j,1},\ldots, {\sf q}i_{j,\#j}\}, \\
        & {\sf number{\_}of{\_}qubits}[{\sf Q}j] \rightarrow \#j
    \end{split}
\end{equation}
that return indices of qubits located in a given qudit and the total number of qubits in a given qudit correspondingly.

The considered `qubit-to-qudit' mapping, in particular Eq.~\eqref{eq:mapping}, implies that the computational basis measurement at the end of the qubit circuit corresponds to a computational basis measurement of qudits, also assumed to be realizable on the qudit-based processor.

\begin{figure}
    \centering
    \includegraphics[width=\linewidth]{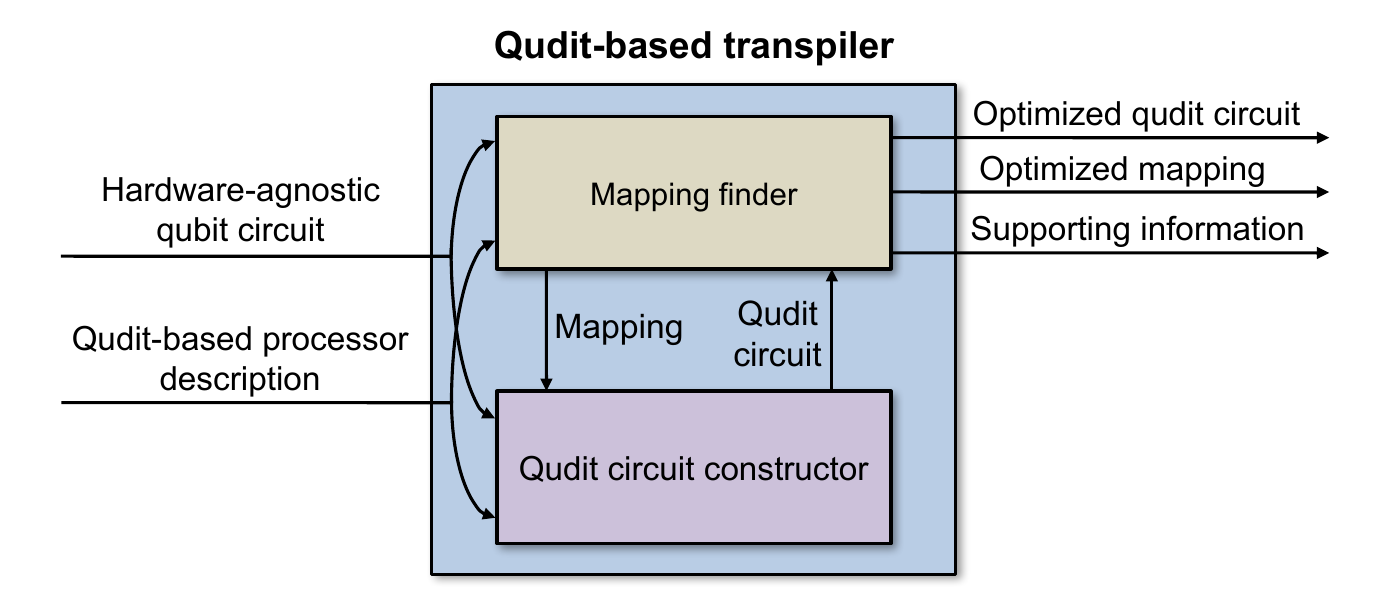}
    \caption{Data transfer scheme in the developed qudit-based transpiler.}
    \label{fig:qudit-tranpiler-scheme}
\end{figure}

The developed qudit transpiler consists of two modules:
(i) the mapping finder and (ii) the qudit circuit constructor (see Fig.~\ref{fig:qudit-tranpiler-scheme}).
Both of them take as input a qudit processor description (values of $m$ and $d$) and a hardware-agnostic qubit-based circuit ${\sf circ}^{\rm qb}$.
The goal of the mapping finder is to search for mapping $\phi_{\rm opt}\in \{ \phi \}$, which minimizes a chosen figure of merit,
while the purpose of the qudit circuit constructor is to generate a qudit circuit ${\sf circ}^{\rm qd}_{\phi}$ that is equivalent to ${\sf circ}^{\rm qb}$ under a mapping $\phi$.
Finally, the qudit-based transpiler outputs the optimized mapping $\phi_{\rm opt}$ and the corresponding circuit ${\sf circ}^{\rm qd}_{\rm opt}:={\sf circ}^{\rm qd}_{\phi_{\rm opt}}$.
The mapping finder can also output some supporting information, which contains, e.g., the exact number of single and two-qudit gates in ${\sf circ}^{\rm qd}_{\rm opt}$ and its comparison with the number of single- and two-qubit gates is in the qubit circuit ${\sf circ}^{\rm qb}_{\rm stand}$ resulted from the standard qubit-based transpilation of ${\sf circ}^{\rm qb}$.
Below we describe the operation of modules in more detail.

\subsection{Mapping finder} \label{sec:mappingfinder}

Here we introduce several approaches of how the optimized qubit-to-qudit mapping can be obtained.
As a figure of merit for a mapping $\phi$ we consider the number of two-qudit gates in ${\sf circ}^{\rm qd}_\phi$.
This choice is motivated by the fact that entangling gates typically represent the main source of fidelity loss.
However, as mentioned before, one can alternative other figures of merits, e.g., circuit depth or fidelity estimations, which can be efficiently calculated given the classical representation of the corresponding qudit-based circuit.

\subsubsection{Finding the optimal mapping with an exhaustive search}
The straightforward way for optimizing qubit-to-qudit mapping is to employ an exhaustive search over all possible mappings $\Phi\equiv\{\phi\}$ of the form Eq.~\eqref{eq:mapping}.
This approach is applicable if the number of available qudits $m$ and their dimension $d$ are reasonably small.

The first step of the exhaustive search is to construct a set of all non-equivalent mappings $\widetilde{\Phi} \subset \Phi$.
Here we call two mappings equivalent if they are different only up to permutations of qubits indices within a particular qudit, or up to permutation of whole sets of qubits' indices belonging to different qudits (and thus definitely provide the same number of entangling gates).
Then, the mapping finder sequentially inputs each $\phi\in\widetilde{\Phi}$ to the qudit circuit constructor to get the corresponding qudit circuit ${\sf circ}^{\rm qd}_{\phi}$.
By comparison of two-qudit gate numbers in ${\sf circ}^{\rm qd}_\phi$ while going through all mappings, the mapping finder chooses the one ($\phi_{\rm opt}$), which provides the smallest number of two-qudit gates.

As we show below, the complexity of generating ${\sf circ}^{\rm qd}_{\phi}$ is linear with respect to the number of gates in the original qubit-based circuit ${\sf circ}^{\rm qb}$, so the possible bottleneck is in the number of mappings in $\widetilde\Phi$.
We note that this issue does not appear in the case of  qutrits ($d=3$), where there is only a single non-equivalent mapping: each qubit ${\sf q}i$ is mapped to a qutrit ${\sf Q}i$.

\begin{figure}
	\centering
	\includegraphics[width=\linewidth]{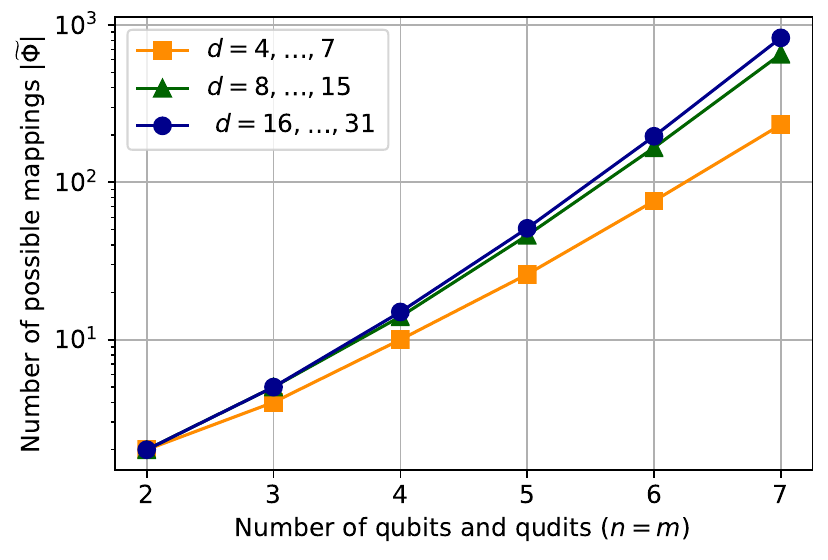}
	\caption{The total number of non-equivalent mapping $|\widetilde{\Phi}|$ depending on the number of qudits $m$ for $d\in\{4,\ldots,31\}$. 
	To maximize the number of nonequivalent mappings, we take the number of qubits $n$ the same as the number of qudits $m$.}
	\label{fig:mappings_num}
\end{figure}

In Fig.~\ref{fig:mappings_num} we show the behavior of the total number of non-equivalent mappings $|\widetilde{\Phi}|$ for different values of $n$ and $d\geq 4$.
Given the fact for $d\leq 31$ and $n \leq 7$, the resulting number of non-equivalent mappings is no more than thousand, it is possible to go through all $\phi\in \widetilde{\Phi}$ within a reasonable time.
We note that in Fig.~\ref{fig:mappings_num} we take the number of qudits $m$ to be equal to the number of qubits $n$, to maximize the number of possible mappings.
Since we deal with a special case of mappings, where each qubit is entirely embedded in a single qudit, the number of mappings for different qudit dimensions taken from a range $d=2^{n'},\ldots,2^{n'+1}-1$ for certain $n'$ is the same.

\begin{figure*}
	\centering
	\includegraphics[width=0.7\linewidth]{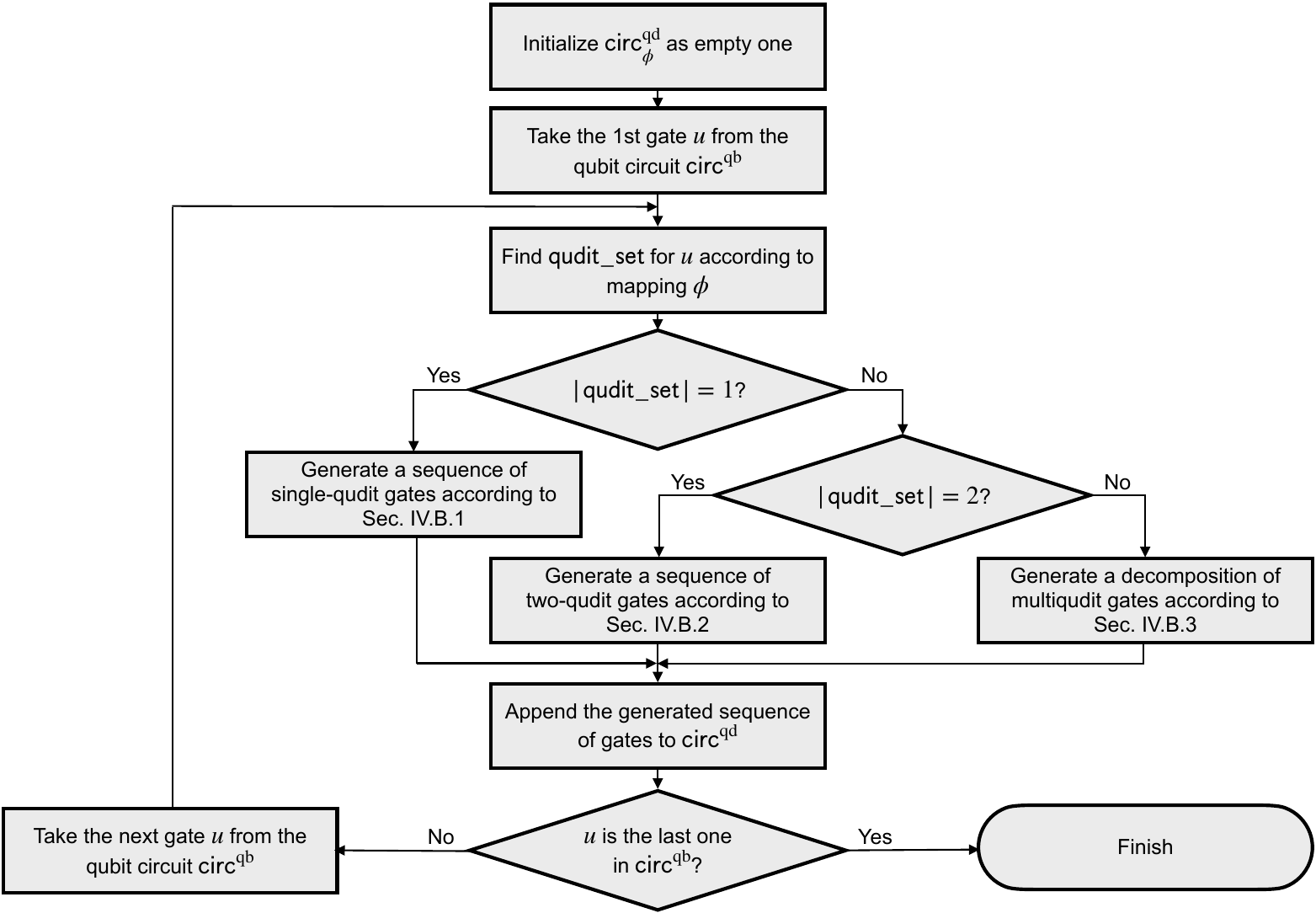}
	\caption{Qudit circuit transpilation algorithm implemented in the qudit-based circuit constructor. 
	Gates from ${\sf circ}^{\rm qb}$ are sequentially classified by the number of qudits in ${\sf qudit{\_}set}$ and then transpiled to qudit-based gates depending on the number of qudits in ${\sf qudit{\_}set}$.}
	\label{fig:transpile_algorithm}
\end{figure*}

\subsubsection{Searching for optimized mappings with polynomial heuristic algorithms}

When the exhaustive search is not applicable, some approximate polynomial methods can be employed.
We emphasize that the problem of finding a mapping from $\Phi$, providing an advantage of using a qudit-based approach compared to the standard qubit-based approach, is much easier than finding the best mapping among all mappings in $\Phi$.
Indeed, in the case of $m\geq n$, the one-to-one mapping $\phi^{(0)}$ definitely provides no more entangling gates compared to the standard qubit-based transpilation, since upper qudit levels are used only for multiqubit gates decomposition. 
{We emphize that in the case of $\phi^{(0)}$, two-qubit entangling ${\sf CZ}$ gates from the input qubit circuit are realized within the qudit-based version exactly in the same way as in the qubit-based one.}
If the input qubit circuit has at least one Toffoli gate, then the number of entangling gates in the corresponding qudit circuit ${\sf circ}^{\rm qd}_{\phi^{(0)}}$ is strictly smaller compared to the one in the standard qubit-based transpilation result ${\sf circ}^{\rm qb}_{\rm stand}$.
Comparing  $\phi^{(0)}$ with some \emph{limited number} candidates from $\Phi$ definitely doesn't make things worse.
In the case of $m<n$, yet $m \lfloor \log_2 d \rfloor \geq n$, the developed qudit-based transpilation method makes it possible to run an $n$-qubit circuit with $m$ $d$-level qudits, which is not possible with $m$ qubits at all.

In the case of $m\geq n$, the choice of candidates from $\Phi$ can be directed by the following observations.
First, it is advisable to consider embedding a pair of two qubits into the space of a single qudit, if there is a relatively huge amount of two-qubit gates connecting these qubits within the given circuit, 
and there is a relatively small amount of gates affecting each of these qubits separately.
Second, it is reasonable to put qubits affected by multiqubit  gates into qudits with free upper levels to use these levels as ancillas for multiqubit  gates decomposition.

One can also consider an iterative ``greedy'' approach of ${\rm poly}(n)$ complexity for finding an optimized mapping,
where a sequential joining of qubits in the space of qudit is considered.
We sketch the idea for the case of $m\geq n$ and $d=4,\ldots,7$ (each qudit can embed no more than two qubits).
Initially, the one-to-one mapping $\phi^{(0)}$ is considered, and the resulting number of entangling gates ${\cal N}_{\rm ent}^{(0)}$ is stored.
At the first step, all $n(n-1)/2$ mappings, where one qudit embeds a qubit pair and $n-2$ other qudits embed remaining $n-2$ qubits are considered.
If the minimal number of entangling gates among these mappings ${\cal N}_{\rm ent}^{(1)}<{\cal N}_{\rm ent}^{(0)}$, then the corresponding mapping $\phi^{(1)}$ with the fewest entangling gates is chosen as a starting point for the next step.
Otherwise, $\phi_{\rm opt}:=\phi^{(0)}$ is the output.
In the second step, $(n-2)(n-3)/2$ mappings with two-qubit pairs (the previously selected pair and a newly tested one) are considered, and so on.
The algorithm proceeds until the number of entangling gates starts to grow, or a mapping with the maximal number $\lfloor n/2 \rfloor$ of qubit pairs is obtained.
Although this algorithm does not guarantee getting the best possible mapping, it provides the resulting number of entangling gates to be no more than the one for a straightforward qubit-based realization, and the maximal number of iterations scales as ${\cal O}(n^3)$.
Given the polynomial complexity (in the number of qubits and number of gates) of the transpilation procedure for a given mapping, we obtain a polynomial complexity of the whole qudit-based transpiler.

It is also worth noting an interesting approach for finding a qubit-to-qudit mapping recently proposed in Ref.~\cite{Mato2023graphs}.
The goal of this algorithm is also to lower the number of non-local operations within the realization of qubit circuits with qudits. 
For this purpose, the authors use a weighted graph representation of a given qubit circuit, where qubit levels represent nodes, graph edges -- local and nonlocal operations, and weights -- the number of corresponding operations.
The authors propose to use an adaption of the $K$-means algorithm to cluster the graph to place edges of the highest weights in distinct clusters.
This clusterization is then interpreted in terms of the qubit-to-qudit mapping.
We note that this algorithm is applied to input qubit circuits \emph{already transpiled down to single- and two-qubit gates}, and therefore does not utilize the full potential of qudits for operating with multiqubit gates, which of the central features of the approach considered in our work.

\subsection{Qudit circuit constructor}\label{sec:qdcircconstructor}

Here we consider in detail how the qudit circuit constructor transpiles qubit circuit ${\sf circ}^{\rm qb}$ to the qudit circuit ${\sf circ}^{\rm qd}_{\phi}$ according to the given qubit-to-qudit mapping $\phi$.
The transpilation process of ${\sf circ}^{\rm qb}$ into ${\sf circ}^{\rm qd}_{\phi}$ is performed in a gate-by-gate principle, shown in Fig.~\ref{fig:transpile_algorithm}.
At the very beginning of the process, ${\sf circ}^{\rm qd}_{\phi}$ is initialized as empty.
Then, for each gate from the qubit circuit ${\sf circ}^{\rm qb}$, the constructor takes a set 
\begin{equation}
	{\sf qubit{\_}set} = \{{\sf q}i_{1},\dots, {\sf q}i_{\kappa}\}.
\end{equation}
of qubits, affected by this gate, and finds the set of corresponding qudits, possessing qubits from ${\sf qubit{\_}set}$:
\begin{equation}\label{eq:qudit_set}
	{\sf qudit{\_}set} = \{
	{\sf qudit\_index}[{\sf q}i_{1}],\ldots, {\sf qudit\_index}[{\sf q}i_{\kappa}] \}.
\end{equation}
We note that ${\sf qudit{\_}set}$ does not contain duplicates of qudit indices. 
Thus, the number of elements in ${\sf qudit{\_}set}$, which we denote by $|{\sf qudit{\_}set}|$, can be less than the number of involved qubits $\kappa$ if several affected qubits are located in the space of the same qudit according to the mapping $\phi$.

The processing of the gate is determined by the value of $|{\sf qudit{\_}set}|$.
If $|{\sf qudit{\_}set}|=1$, then the qubit gate under processing can be realized on the qudit processor as a sequence of single-qudit gates (see subsection~\ref{sec:single-qd-case}). 
In the case of $|{\sf qudit{\_}set}|>1$, two-qudit gates become necessary.
It is convenient to distinguish the case of $|{\sf qudit{\_}set}|=2$ and the case of $|{\sf qudit{\_}set}|\geq 3$ that we describe in detail in sections~\ref{sec:two-qd-case} and~\ref{sec:mlt-qd-case}, correspondingly.
For all these cases, we obtain a sequence of qudit gates that implement the processed qubit gate.
This sequence is added to the end of ${\sf circ}^{\rm qd}_{\phi}$, and then the procedure is repeated for the next gate from ${\sf circ}^{\rm qb}$ until all gates have been processed.
Below we describe the exact decomposition of qubit gates into the sequence of qudit gates for all possible values of $|{\sf qudit{\_}set}|$.

\subsubsection{Single-qudit case}\label{sec:single-qd-case}

The case of $|{\sf qudit{\_}set}|=1$ can appear in two situations: 
(i) the processed gate is a single-qubit one ($\kappa=1$), and (ii) the processed gate in a multi-qubit ($\kappa\geq2$) with all affected qubits being located in the same qudit according to the mapping $\phi$.

\begin{figure}
	\centering
	\includegraphics[width=0.9\linewidth]{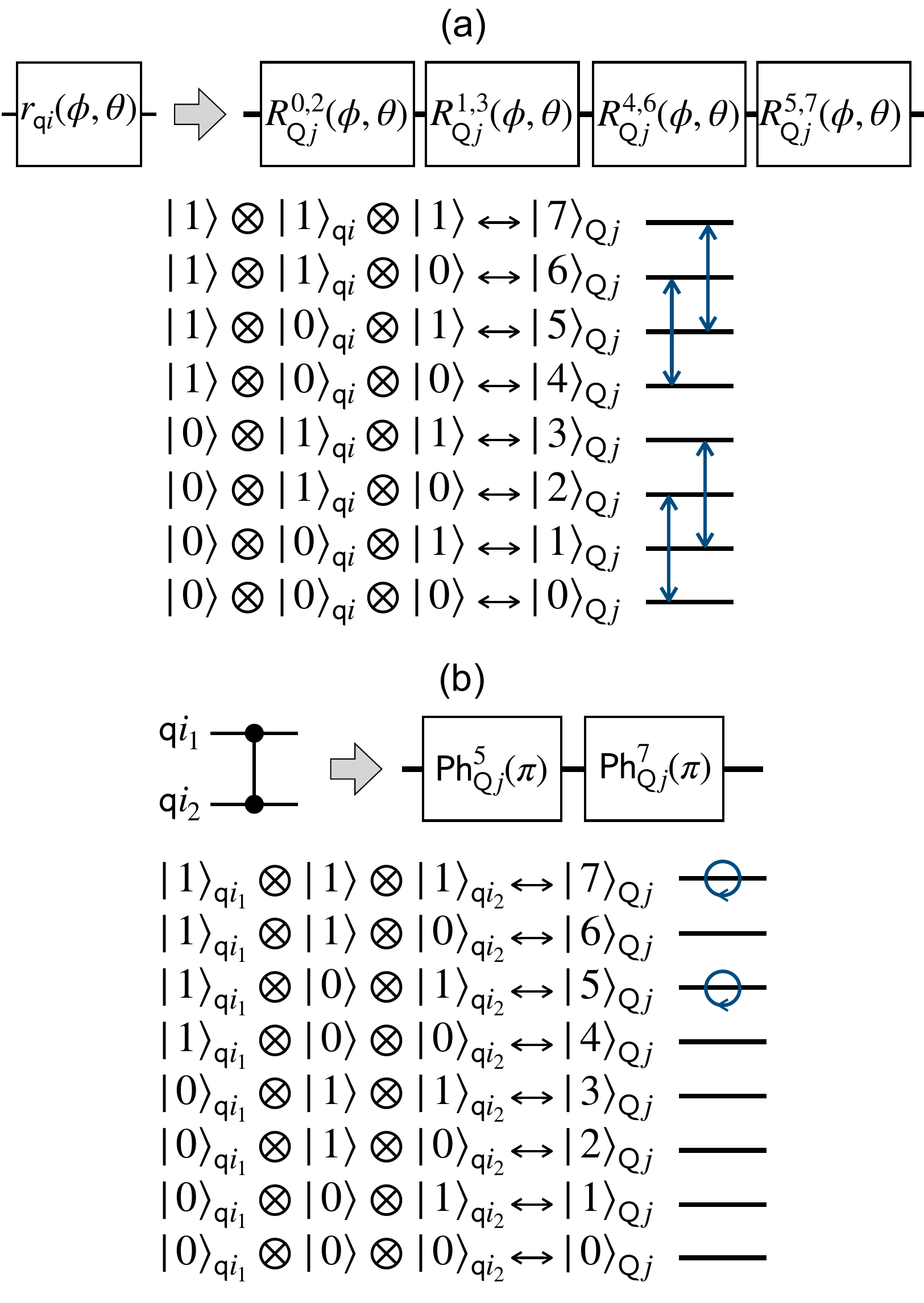}
	\caption{(a) Qudit-based realization of a single-qubit gate $r_{{\sf q}i}(\varphi,\theta)$ in the case where ${\sf q}i$ is embedded into $8$-level qudit ${\sf Q}j$ at the 2nd position (${\sf Q}j$ contains three qubits in total).
	The involved transitions within the $8$-level qudit ${\sf Q}j$ are shown.
	(b) Qudit-based realization of a two-qubit gate ${\sf CZ}_{{\sf q}i_1,{\sf q}i_2}$ in the case where the affected qubits ${\sf q}i_1$ and ${\sf q}i_2$ are embedded into the same qudit ${\sf Q}j$ at the 1st and 3rd positions, correspondingly.
	The levels acquiring the phase factor of -1 are shown.}
	\label{fig:single-qd-case}
\end{figure}

First, let us consider the case of a single-qubit gate, acting on a qubit ${\sf q}{i_1}$.
Let ${\sf Q}j={\sf qudit\_index}[{\sf q}{i_1}]$, 
${\sf pos}={\sf position\_in\_qudit}[{\sf q}{i_1}]$, and $\#j={\sf number\_of\_qubits}[{\sf Q}j]$.
Remind that in our realization, the only type of single-qubit gates is rotation $r_{{\sf q}i}(\varphi,\theta)$, defined in Eq.~\eqref{eq:qubit_rotation}.
To implement this unitary in the qudit's space, we need to consider a tensor product of a $2\times 2$ unitary acting in a subspace of affected qubits with an identity operation acting in the remaining space of a qudit. 
The resulting correspondence between the qubit gate and a sequence of qudit gates is given by
\begin{equation} \label{eq:singlequbitgatemapping}
	r_{{\sf q}i}(\varphi,\theta) \to
	\prod\limits_{(\alpha, \beta)} R_{{\sf Q}i}^{\alpha,\beta}(\varphi,\theta),
\end{equation}
where the product is made over all possible pair levels $(\alpha,\beta)$ satisfying the following condition:
\begin{equation} \label{eq:singleqbitbins}
	\begin{split}
		\bin(\alpha)&=x_1\dots x_{{\sf pos}-1} 0~ x_{{\sf pos}+1}\dots x_{\#j}, \\
		\bin(\beta)&=x_1\dots x_{{\sf pos}-1} 1~ x_{{\sf pos}+1}\dots x_{\#j}.
	\end{split}
\end{equation}
Here $\bin(\alpha)$ and $\bin(\beta)$ are $\#j$-length binary representation of $\alpha$ and $\beta$, correspondingly 
(let us remind that $\alpha,\beta\in\{0,1,\ldots,d-1\}$ and $\#j \leq \log_2 d$), and $x_1\dots x_{{\sf pos}-1}x_{{\sf pos}+1}\dots x_{\#j}$ are all possible bitstrings of length $\#j-1$.
The sequence of qudit unitaries, given by Eqs.~\eqref{eq:singlequbitgatemapping} and~\eqref{eq:singleqbitbins}, 
is in agreement with the employed structure of qubit-to-qudit mappings shown in Eq.~\eqref{eq:binarymapping} [see also Fig.~\ref{fig:single-qd-case}(a) for an intuitive explanation].

\begin{figure}
	\centering
	\includegraphics[width=0.9\linewidth]{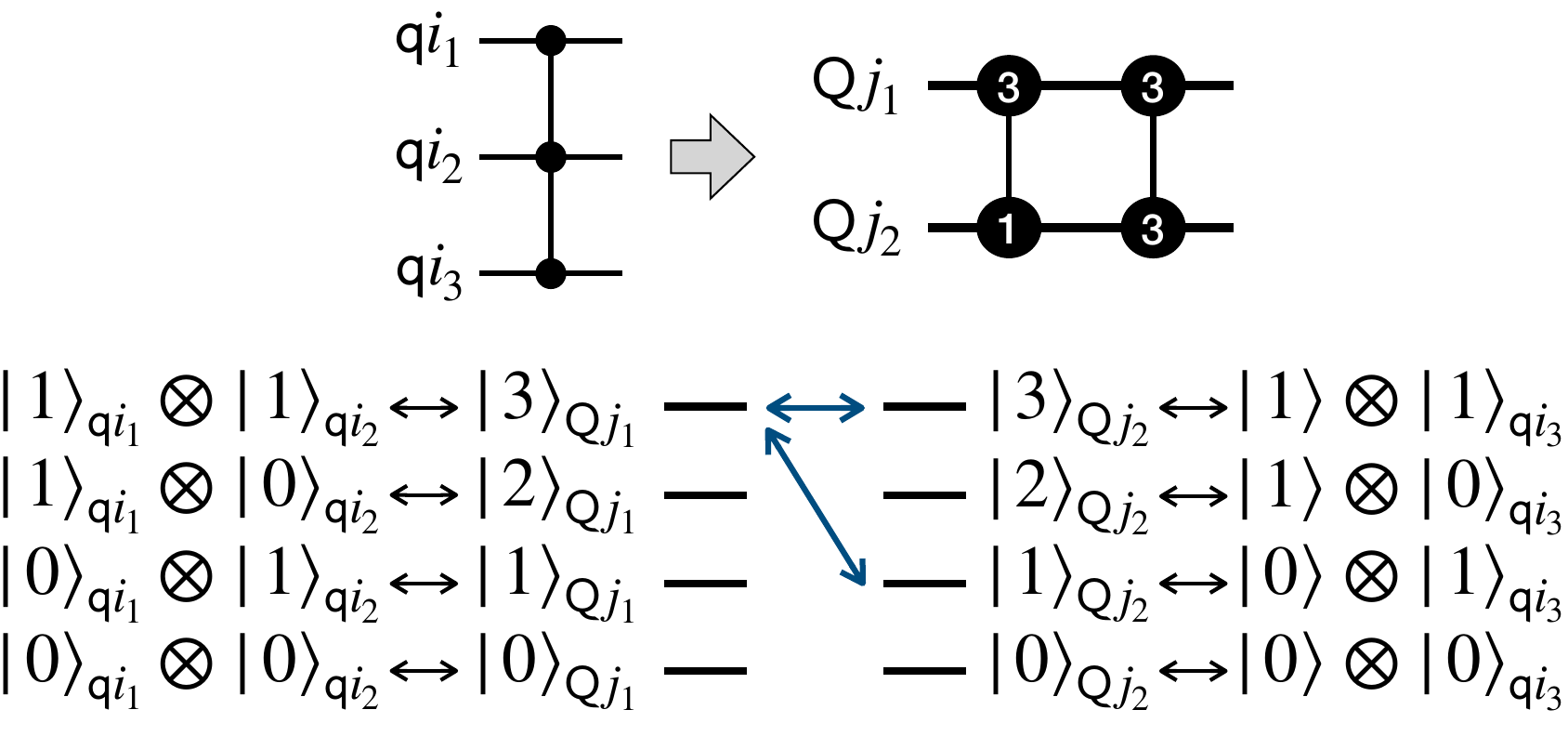}
	\caption{
	Qudit-based realization of a three-qubit gate ${\sf CZ}_{{\sf q}i_1,{\sf q}i_2,{\sf q}i_3}$ in the case where ${\sf q}i_1$ and ${\sf q}i_2$ are embedded into ${\sf Q}j_1$ and ${\sf q}i_3$ together 
	with some other qubit are embedded into ${\sf Q}j_2$ correspondingly (${\sf Q}j_1$ and ${\sf Q}j_2$ are 4-level systems).}
	\label{fig:two-qd-case}
\end{figure}

Let us then consider the case of a multi-qubit gate ${\sf CZ}_{{\sf q}i_1,\ldots, {\sf q}i_\kappa}$, where all qubits ${\sf q}i_1,\ldots, {\sf q}i_\kappa$ are located in the same qudit ${\sf Q}j$.
Let ${\sf pos}_{i_k}={\sf position\_in\_qudit}[{\sf q}{i_k}]$ for $k=1,\ldots,\kappa$.
Then the desired gate ${\sf CZ}_{{\sf q}i_1,\ldots, {\sf q}i_\kappa}$ can be realized with the following sequence of single-qudit phase gates:
\begin{equation} 
    {\sf CZ}_{{\sf q}i_1,\ldots, {\sf q}i_\kappa} \to
   \prod\limits_{\alpha}{\rm Ph}^{\alpha}_{{\sf Q}j}(\pi),\\
\end{equation}
where qudit levels $\alpha$ satisfy the following condition:
\begin{equation}
       \bin(\alpha)[{\sf pos}_{i_k}] = 1~\text{for~each}~k=1,\dots, \kappa
\end{equation}
(here $\bin(\alpha)[{\sf pos}_{i_k}]$ stands for ${\sf pos}_{i_k}$th bit in a $\#j$-length binary representation of $\alpha$).
As in the case of a single-qubit gate, the resulting sequence provides a proper transformation at the required qudit levels [see Fig.~\ref{fig:single-qd-case}(b)].
Clearly, the processing of the single-qudit case is of ${\cal O}(1)$ space and time complexity on a classical computer.

\subsubsection{Two-qudit case}\label{sec:two-qd-case}

Here we consider the case where qubits, which are involved in certain multi-qubit gate ${\sf CZ}_{{\sf q}i_1,\ldots,{\sf q}i_\kappa}$, are located (according to $\phi$) in two different qudits, namely ${\sf Q}j_1$ and ${\sf Q}j_2$.
Let $\{{\sf pos}_1^{l}\}$ and $\{{\sf pos}_2^{l'}\}$ be positions, according to ${\sf position\_in\_qudit}[\cdot]$, of qubits located at ${\sf Q}j_1$ and ${\sf Q}j_2$, correspondingly.

As in the case of a single-qubit gate, the resulting transformation performed in the space of two qudits is obtained as a tensor product of the unitary corresponding to 
${\sf CZ}_{{\sf q}i_1,\ldots,{\sf q}i_\kappa}$ in the proper subspace of the two-qudits space and identity operator in the remaining subspace.
This operation reads
\begin{equation}  \label{eq::for_type_C}
	{\sf CZ}_{{\sf q}i_1, \dots, {\sf q}i_\kappa} \to \prod\limits_{(\alpha,\beta)} {\sf CZ}^{\alpha, \beta}_{{\sf Q}j_1, {\sf Q}j_2},
\end{equation}
where pairs of levels $(\alpha,\beta)$ are all possible admissible pairs satisfying the condition 
\begin{equation} \label{eq:conditionforgeneralcz}
	\bin(\alpha)[{\sf pos}^\ell_{1}] = 1, \quad \bin(\beta)[{\sf pos}^{{\ell'}}_{2}] = 1.
\end{equation}
According to Eqs.~\eqref{eq::for_type_C} and~\eqref{eq:conditionforgeneralcz}, the number of ${\sf CZ}^{\alpha, \beta}_{{\sf Q}j_1, {\sf Q}j_2}$ gates in the resulting sequence is determined by the number of qubits located in qudits ${\sf Q}j_1$ and ${\sf Q}j_2$ and {\it not} involved by ${\sf CZ}_{{\sf q}i_1, \dots, {\sf q}i_\kappa}$ (see an example for two-qudit case in Fig.~\ref{fig:two-qd-case}).
Each unused qubit doubles the number of pairs $(\alpha,\beta)$ satisfying Eq.~\eqref{eq:conditionforgeneralcz}, and so the resulting number of two-qudit gates is given by
\begin{equation} \label{eq:penalty}
	2^{\#j_1+\#j_2-\kappa},
\end{equation}
where $\#j_1$ and $\#j_2$ are the number of qubits in ${\sf Q}j_1$ and ${\sf Q}j_2$, correspondingly.

Eq.~\eqref{eq:penalty} captures an intuition behind qudit ${\sf CZ}$ gate.
On the one hand, implementation of, e.g., two-qubit gates ($\kappa=2$) in the case where qudits ${\sf Q}j_1$ and ${\sf Q}j_2$ 
contain other qubits ($\#j_1+\#j_2>2$) costs more two-body ${\sf CZ}$-type interactions than in the case of direct qubit-based realization ($2^{\#j_1+\#j_2-1}$ compared to 1).
{Recall, however, that in the case of the one-to-one mapping $\phi^{(0)}$, $\#j_1=\#j_2=1$ and $\kappa=2$, so there in no overhead in the number of entnagling gates.}
On the other hand, in the case of multi-qubit gates with $\kappa>2$, the resulting number of two-body ${\sf CZ}$-type interactions can become smaller compared to the one 
obtained from known multi-qubit gates decompositions into single-qubit and two-qubit gates (see e.g. Ref.~\cite{Barenco1995}).
We also remind that in the case where all qubits affected by ${\sf CZ}_{{\sf q}i_1, \dots, {\sf q}i_\kappa}$ fall into the same qudit, there is no need for two-body interactions at all.

As in the single-qudit case, the described processing is of ${\cal O}(1)$ space and time complexity on a classical computer.
{We also note that for other types of two-qudit interactions, different from ${\sf CZ}^{\alpha, \beta}_{{\sf Q}j_1, {\sf Q}j_2}$, Eqs.~\eqref{eq::for_type_C} and~\eqref{eq:penalty} have to be modified.  
For example, as shown in~\cite{nikolaeva2023universal}, in a trapped-ion platform with native parametric two-qudit M$\o$lmer-S$\o$rensen gate, to implement two-qubit ${\sf CZ}$ gates ($\kappa=2$) in the case, where qudits ${\sf Q}j_1$ and ${\sf Q}j_2$ contain two qubits each, a single two-qudit M$\o$lmer-S$\o$rensen gate with increased value of effective rotation angle, compared to the rotation angle used within the qubit-based verstion, is needed.}

\subsubsection{Multi-qudit gate case}\label{sec:mlt-qd-case}

Here we describe the most complicated case, where qubits affected by the gate ${\sf CZ}_{{\sf q}i_1, \dots, {\sf q}i_\kappa}$, fall into more than two qudits.
To make the decomposition description more clear, let us introduce new notations.
For each qudit ${\sf Q}j$, we define states $\zero_{{\sf Q}j}\equiv \ket{0}_{{\sf Q}j}$ and $\one_{{\sf Q}j}\equiv \ket{d^2-1}_{{\sf Q}j}$  
that corresponds to multi-qubit states $\ket{0\dots 0}_{{\sf q}i_{j,1},\ldots,{\sf q}i_{j,\# j}}$ and $\ket{1\dots 1}_{{\sf q}i_{j,1},\ldots,{\sf q}i_{j,\# j}}$, correspondingly, 
with respect to the considered mapping $\phi$ (remind that ${\sf q}i_{j,1},\ldots,{\sf q}i_{j,\# j}$ denote labels of qubits located in the space of qudit ${\sf Q}j$).
If qudit dimension $d>2^{\#j}$, we also define an ancillary state $\ket{{\rm a}}_{{\sf Q}j}\equiv \ket{2^{\#j}}_{{\sf Q}j}$ that is beyond the qubits' subspace in the space of ${\sf Q}j$, 
and a flag, indicating whether the ancillary level is available in this qudit:
\begin{equation}
	{\sf ancilla}[{\sf Q}j] = 
	\begin{cases}
		{\sf True}, &\text{if~} d \geq 2^{\#j},\\
		{\sf False}, &\text{otherwise}.
	\end{cases}
\end{equation}

Let ${\sf qudit\_set}$ be a set of qudit indices involved in the realization of ${\sf CZ}_{{\sf q}i_1,\ldots, {\sf q}i_\kappa}$, see Eq.~\eqref{eq:qudit_set}.
We assign each qudit from ${\sf qudit{\_}set}$ to one of three possible types labeled as ${\cal A}$, ${\cal B}$, or ${\cal C}$.

We say that ${\sf Q}j$ belongs to type ${\cal A}$, if all qubits, located in this qudit, are affected by ${\sf CZ}_{{\sf q}i_1, \dots, {\sf q}i_\kappa}$ and ${\sf Q}j$ has no ancillary level, that is
\begin{equation}
	\left({\sf indices{\_}of{\_}qubits}({\sf Q}j) \subset{\sf qubit{\_}set}\right)\wedge  \overline{{\sf ancilla}[{\sf Q}j]},
\end{equation}
where $\overline{{\sf ancilla}[{\sf Q}j]}$ stands for ${\sf ancilla}[{\sf Q}j]={\sf False}$.
If all qubits located in qudit ${\sf Q}j$ are involved in the decomposed qubit gate and ${\sf Q}j$ has ancillary level, that is
\begin{equation}
	\left({\sf indices{\_}of{\_}qubits}[{\sf Q}j] \subset {\sf qubit{\_}set}\right) \wedge  {\sf ancilla}[{\sf Q}j],
\end{equation}
then we say that qudit ${\sf Q}j$ belongs to type $\cal B$.
Qudit ${\sf Q}j\in {\sf qudit\_set}$ belongs to type ${\cal C}$ if there is at least one qubit located in ${\sf Q}j$ but not affected by ${\sf CZ}_{{\sf q}i_1,\ldots, {\sf q}i_\kappa}$:
\begin{equation}
	({\sf Q}j \in {\sf qudit\_set}) \wedge ({\sf indices{\_}of{\_}qubits}[{\sf Q}j] / {\sf qubit{\_}set} \neq \varnothing).  
\end{equation}
We denote the number of qudits of types ${\cal A}$, ${\cal B}$, and ${\cal C}$ as $|{\cal A}|$,  $|{\cal B}|$, and $|{\cal C}|$, correspondingly.

\begin{figure*}
	\centering
	\includegraphics[width=\linewidth]{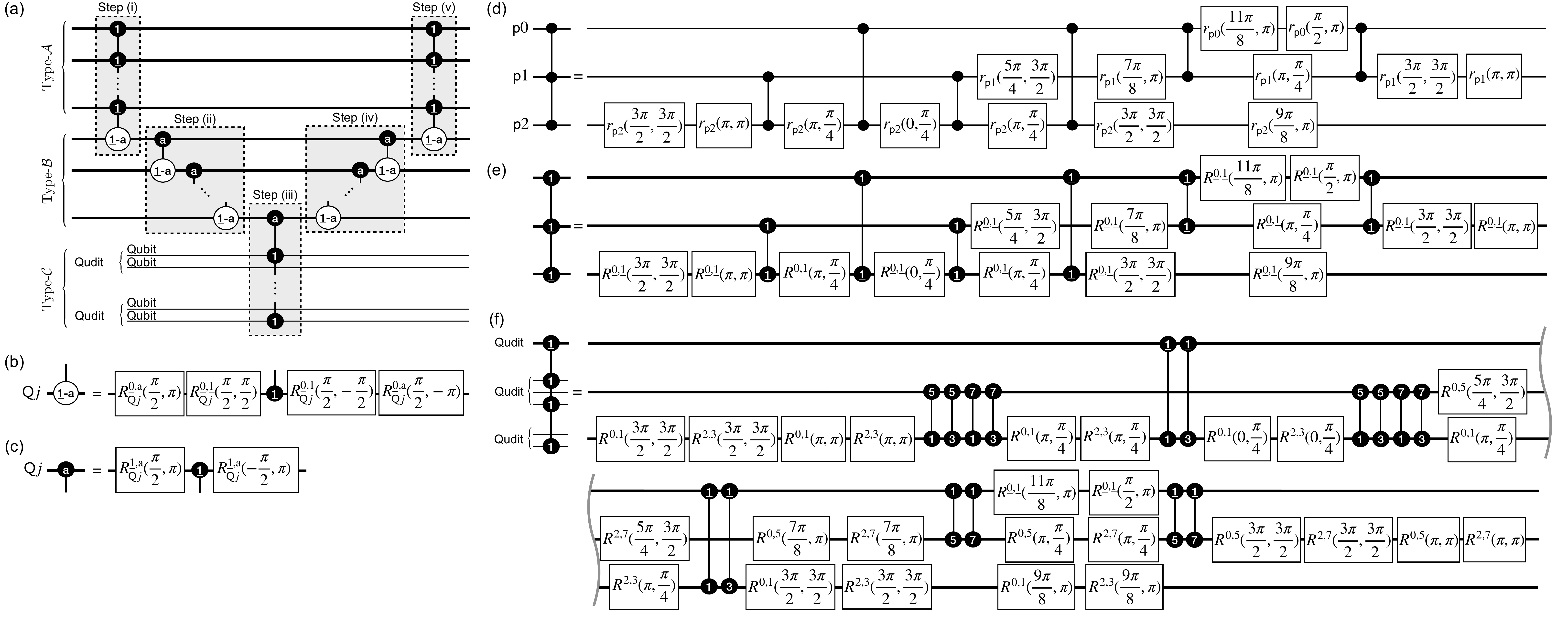}
	\caption{(a) Intermediate qudit circuit ${\sf circ}^{\text{qd-int}}_{\phi}$ used for the decomposition of ${\sf CZ}_{{\sf q}i_1,\ldots, {\sf q}i_\kappa}$ with $|{\sf qudit\_set}|\geq 3$.
	Each gate of ${\sf circ}^{\text{qd-int}}_{\phi}$ is then transpiled down to native single-qudit and two-qudit gates.
	(b) The scheme of transforming an inversion operation between levels $\ket{\underline{1}}$ and $\ket{\rm a}$ to the phase accruing operation at level $\ket{\underline{1}}$ via single-qudit operations. 
	(c) The scheme of transforming a control at level $\ket{\rm a}$ to the control at level $\ket{\underline{1}}$ via single-qudit operations.
	(d) An example of the qubit-based ancilla-free decomposition of three-qubit ${\sf CZ}_{{\sf p}0,{\sf p}1,{\sf p}2}$ gate according Ref.~\cite{Barenco1995} that can be used as a template to decompose multi-qudit $\sf CZ$ type gates at Steps (i), (iii), and (v).
	(e) Decomposition of a three-qudit gate used in Steps (i) and (v) to single-qudit and two-qudit gates according to the template, shown in (d).
	(f) Decomposition of a three-qudit gate used in Step (iii) to single-qudit and two-qudit gates according to the template, shown in (d).}
	\label{fig:multi-qudit-gate}
\end{figure*}

The transpilation of ${\sf CZ}_{{\sf q}i_1,\ldots, {\sf q}i_\kappa}$ is performed by, first, constructing an intermediate qudit circuit ${\sf circ}^{\text{qd-int}}_{\phi}$ presented in Fig.~\ref{fig:multi-qudit-gate}(a), 
and then decomposing ${\sf circ}^{\text{qd-int}}_{\phi}$ down to single-qudit and two-qudit gates.

We then split ${\sf circ}^{\text{qd-int}}_{\phi}$ into five steps: (i) multi-qudit controlled gate with controls on type $\cal A$ qudits and target on the first qudit of type $\cal B$;
(ii) down-step ladder-like sequence of two-qudit gates on all qudits of type $\cal B$;
(iii) multi-qudit gate acting on the last qudit of type $\cal B$ and all qudits of type $\cal C$;
(iv) up-step ladder-like sequence that is the uncomputation of step (ii);
(v) the uncomputation of step (i).

The general idea of this structure is as follows.
According to ${\sf CZ}_{{\sf q}i_1,\ldots, {\sf q}i_\kappa}$, we have to add a phase factor of $-1$ to all basis states of involved qudits such that all corresponding qubits ${\sf q}i_1,\ldots, {\sf q}i_\kappa$, embedded in these qudits, are in the state $1$.
We note that for qudits of types $\cal A$ and $\cal B$, there is a single level that is a candidate for acquiring the phase factor, namely $\one$.
For type $\cal C$ qudits, the situation is different.
Since the presence of unaffected qubits, the phase factor is acquired or not acquired (depending on the state of other involved qudits), to several levels of type ${\cal C}$ qudit, namely to all levels $\alpha$ satisfying the condition
\begin{equation}
	\bin(\alpha)[{\sf pos}]=1,
\end{equation}
where ${\sf pos}$ values correspond to positions of affected qubits in the considered qudit.
Roughly speaking, we need to add a phase factor of $-1$ to a computational basis state of involved qudits, if all type-$\cal A$ and type-$\cal B$ qudits are in the state $\one$, and all qubits encoded in type-$\cal C$ qudits are in the state $\ket{1}$.

The important feature of type $\cal B$-qudits is that they possess an ancillary level that can be employed for storing temporary information within the gate decomposition.
We use an ancillary level of type $\cal B$ qudit for storing a `flag' whether this qudit and all `previous' qudits are in the proper state for acquiring the phase factor: this is the way how the ladder type sequences [parts (ii) and (iv)] appear in our construction.
As we discuss further, the operation with qudits of type $\cal A$ and $\cal C$ is based on standard schemes of reconstructing multi-qubit controlled gates down to single-qubit and two-qubit gates.
At the same time, the implementation of a two-qubit operation for qubits in qudits, possessing other uninvolved qubits (qudits of type $\cal C$), results in overhead in the number two-qudit operation (as we discussed in Section~\ref{sec:two-qd-case}).
In order to avoid the doubling of this overhead in the uncomputation, we put the operation with $\cal C$ type qudits in the middle of our circuit.

However, it is a possibly realizable situation when the number of two-qudit gates required for processing $\cal C$ type qudits is lower than the number of two-qudit gates for processing $\cal A$ type qudits.
In this case, it is preferable to swap $\cal A$ and $\cal C$-type qudits in the structure of ${\sf circ}^{\text{qd-int}}_{\phi}$.
In order to simplify our description, next we consider the processing of the original circuit presented in Fig.~\ref{fig:multi-qudit-gate}(a).
The possible improvement related to swapping $\cal A$ and $\cal C$ qudits in the structure of ${\sf circ}^{\text{qd-int}}_{\phi}$ is discussed in Appendix~\ref{sec:alt_int_scheme}.
We also consider modifications of the described scheme in cases where one or two types of qudits are missing (e.g. $|{\cal B}|=0$) in Appendix~\ref{sec:special_cases}.

Below we consider the decomposition of each of the described groups of gates to the set of basic single-qudit and two-qudit gates for our transpiler. 

\paragraph{Processing the multi-controlled gate of Step (i).} \label{sec:type-a-cz}

The idea of its decomposition is as follows.
First, by employing single-qudit rotations $R_{{\sf Q}j}^{\underline{0},\underline{1}}(\phi,\theta)$ and $R_{{\sf Q}j}^{\underline{0},{\rm a}}(\phi,\theta)$ where ${\sf Q}j$ is an affected type-${\cal B}$ qudit, and notations $\underline{0}\equiv 0$, $\underline{1}\equiv d^2-1$, and ${\rm a}\equiv 2^{\#j}$ are used, we turn the desired multi-qudit gate into the gate of $\sf CZ$ type [see Fig.~\ref{fig:multi-qudit-gate}(b)].
Namely, it adds the phase factor $-1$ to the state $\one\otimes\ldots\otimes\one$ of the affected qudits and leaves the remaining states unchanged.

Then we take an ancilla-free decomposition of $(|{\cal A}|+1)$-qubit gate ${\sf CZ}_{{\sf p}0,\ldots,{\sf p}(|{\cal A}|)}$ gate, acting on abstract qubits ${\sf p}0,\ldots,{\sf p}|{\cal A}|$, to single-qubit gates $r_{{\sf p}k}(\phi,\theta)$ and two-qubit gates ${\sf CZ}_{{\sf p}k_1 {\sf p}k_2}$.
This decomposition is used as a `template' for reconstructing the desired multi-qudit gate into a single-qudit and two-qudit gate. 
As an example, one can take a standard decomposition from Ref.~\cite{Barenco1995} shown in Fig.~\ref{fig:multi-qudit-gate}(d).
Next we turn each single-qubit gate $r_{{\sf p}k}(\phi,\theta)$ into a single-qudit gate $R^{\underline{0},\underline{1}}_{{\sf Q}j}(\phi,\theta)$, and each ${\sf CZ}_{{\sf p}k_1 {\sf p}k_2}$ into ${\sf CZ}^{\underline{1},\underline{1}}_{{\sf Q}j_1 {\sf Q}j_2}$, where the correspondence between qubits ${\sf p}0,\ldots,{\sf p}|{\cal A}|$ and affected qudits is realized via a straightforward ordering (see Fig.~\ref{fig:multi-qudit-gate} (e)).
One can see that this construction provides the realization of ${\sf CZ}$ operation in the space spanned by a tensor product of states $\zero$ and $\one$ of affected qudits.

We note that it is possible a situation, where the taken qubit-based decomposition of ${\sf CZ}_{{\sf p}0,\ldots,{\sf p}|{\cal A}|}$ realizes the gate up to a global phase.
In our case, where we embed this decomposition into qudit space, the global phase turns into a relative one between the ${\sf CZ}$ 
operation in the subspace spanned by a tensor product of $\zero$ and $\one$ states and the identity operation in the remaining subspace.
However, this relative phase is removed in the uncomputation Step (v), given that all operations in the uncomputation are Hermitian conjugates of ones in Step (i).

\paragraph{Processing the ladder-like sequence on type-$\cal B$ qudits of Step (ii).}\label{sec:type-b-cz}

Remind that each type-$\cal B$ qudit has an ancillary level $\ket{\rm a}$ that we use to store the information about whether `previous' (according to the ordering in Fig.~\ref{fig:multi-qudit-gate}(a)) qudits are in the state $\one$.
The idea behind employed gates in the ladder-like sequence of Step (ii) is quite straightforward: Each gate turns the state of the target qudit from $\one$ to $\ket{\rm a}$ if and only if the control qudit is in the state $\ket{\rm a}$.
One can see that by realizing the sequence of Step (ii), following Step (i), the last type-$\cal B$ qudit appears in the state $\ket{\rm a}$ if and only if all type-$\cal A$ and type-$\cal B$ qudits were initially in the state $\one$.
The decomposition employed in Step (ii) two-qudit gates using native qudit gates can be performed according to schemes of Fig.~\ref{fig:multi-qudit-gate}(b) and (c).

\paragraph{Processing the multi-controlled gate of Step (iii).}\label{sec:type-c-cz}

The goal of the considered gate is to acquire the phase factor of $-1$ to the input state if the last type-${\cal B}$ qudit is in the ancillary state $\ket{\rm a}$, 
and type-$\cal C$ qudits are in a such state that all qubits embedded in the type-$\cal C$ qudits and affected by the decomposed gate are in the state $\ket{1}$.
As has been mentioned, the important point of type-$\cal C$ qudits is that they also contain unaffected qubits, which results in the fact that the gate of Step (iii) has to acquire the phase factor of $-1$ to several computational basis states.
Specifically, the phase factor of $-1$ has to be acquired to $2^{\#{\rm unaff}}$ states, where $\#{\rm unaff}$ is a total number of unaffected qubits in type-$\cal C$ qudits.
The intuition behind this fact is exactly the same as behind Eq.~\eqref{eq:penalty} of the required number of two-qudit gates for realizing an operation between qubits embedded in these qudits.

The idea of decomposing the gate of Step (iii) is very similar to the one of decomposing the gate at Step (i).
First, we turn the gate at Step (iii) to a multi-controlled gate of ${\sf CZ}$ type by adding $R_{{\sf Q}j}^{\underline{1}, {\rm a}}(\pi/2,\pi)$ and $R_{{\sf Q}j}^{\underline{1}, {\rm a}}(\pi/2,-\pi)$ rotations on the last qudit ${\sf Q}j$ of type ${\cal B}$. 
Then we take an ancilla-free decomposition of $(|{\cal C}|+1)$-qubit ${\sf CZ}_{{\sf p}0,\ldots,{\sf p}(|{\cal C}|)}$ gate 
(acting on virtual qubits ${\sf p}0,\ldots,{\sf p}|{\cal C}|$) to single-qubit rotations $r_{{\sf p}k}(\phi,\theta)$ and two-qubit ${\sf CZ}_{{\sf p}k_1{\sf p}k_2}$ gates [see Fig.~\ref{fig:two-qd-case}(f)].

Each single-qubit gate $r_{{\sf p}k}(\phi,\theta)$ is transformed into the sequence of single-qudit gates
\begin{equation}
    \prod\limits_{(\alpha,\beta)}R^{\alpha,\beta}_{{\sf Q}j}(\varphi,\theta),
\end{equation}
where ${\sf Q}j$ is a qudit corresponding to ${\sf p}k$ according to the straightforward ordering, and $(\alpha, \beta)$ are all appropriate level pairs satisfying the condition
\begin{equation}
    \begin{split}
        &\bin(\alpha)[{\sf pos}] = 0, \bin(\beta)[{\sf pos}] = 1, \text{~for~} {\sf pos}\in \{{\sf pos}_\ell\}; \\
        &\bin(\alpha)[{\sf pos}] = \bin(\beta)[{\sf pos}], \text{~for~} {\sf pos}\notin \{{\sf pos}_\ell\}, \\
    \end{split}
\end{equation}
and $\{{\sf pos}_\ell\}$ is a set of position of qubits embedded in ${\sf Q}j$ and affected by the multi-qudit gate of Step (iii). 

Two-qubit gates ${\sf CZ}_{{\sf p}k_1{\sf p}k_2}$ are transformed according to Sec.~\ref{sec:two-qd-case}.
Namely, we take qudits ${\sf Q}j_1$ and ${\sf Q}j_2$ corresponding to ${\sf p}k_1$ and ${\sf p}k_2$, chose the set of qubits ${\sf q}i_1',\ldots {\sf q}i_\kappa'$ that are embedded in ${\sf Q}j_1$ and ${\sf Q}j_2$ and are affected by the multi-qudit gate of Step (iii), and transform ${\sf CZ}_{{\sf q}i_1',\ldots {\sf q}i_\kappa'}$ according to~\eqref{eq::for_type_C} and~\eqref{eq:conditionforgeneralcz}. 

In contrast to the decomposition of the multi-controlled gate of Step (i), here we should also take into account the global phase factor that may appear from the qubit-based decomposition.
Though it is insignificant in the qubit case, when we use this decomposition for qubit embedded into qudits, the phase turns from the global to the relative one 
(this is the relative phase between the ${\sf CZ}$ operation in the subspace of affected qubits, and identity operation in the remaining subspace).
To compensate for this phase explicitly, we add a phase single-qubit gate
\begin{equation}
	{\sf Ph}_{{\sf p}0}(\gamma)=\begin{bmatrix}
		e^{\imath \gamma} & 0\\
		0 & e^{\imath \gamma}
	\end{bmatrix}
\end{equation}
to the qubit ${\sf p}0$.
The value of $\gamma$ is chosen to make the whole sequence of gates, applied to ${\sf p}0,\ldots,{\sf p}|{\cal C}|$, to realize ${\sf CZ}_{{\sf p}0,\ldots,{\sf p}|{\cal C}|}$ without any global phase.

From the viewpoint of the qudit circuit, ${\sf p}0$ corresponds to the type-$\cal B$ qudit ${{\sf Q}j}$ involved in the multi-controlled gate of Step (iii).
The qubit phase gate ${\sf Ph}_{{\sf p}0}(\gamma)$ transforms into
\begin{equation} 
	{\sf Ph}_{{\sf Q}j}^{\underline{0}}(\gamma){\sf Ph}_{{\sf Q}j}^{\underline{1}}(\gamma).
\end{equation}

\begin{figure*}[]
\center{\includegraphics[width=\linewidth]{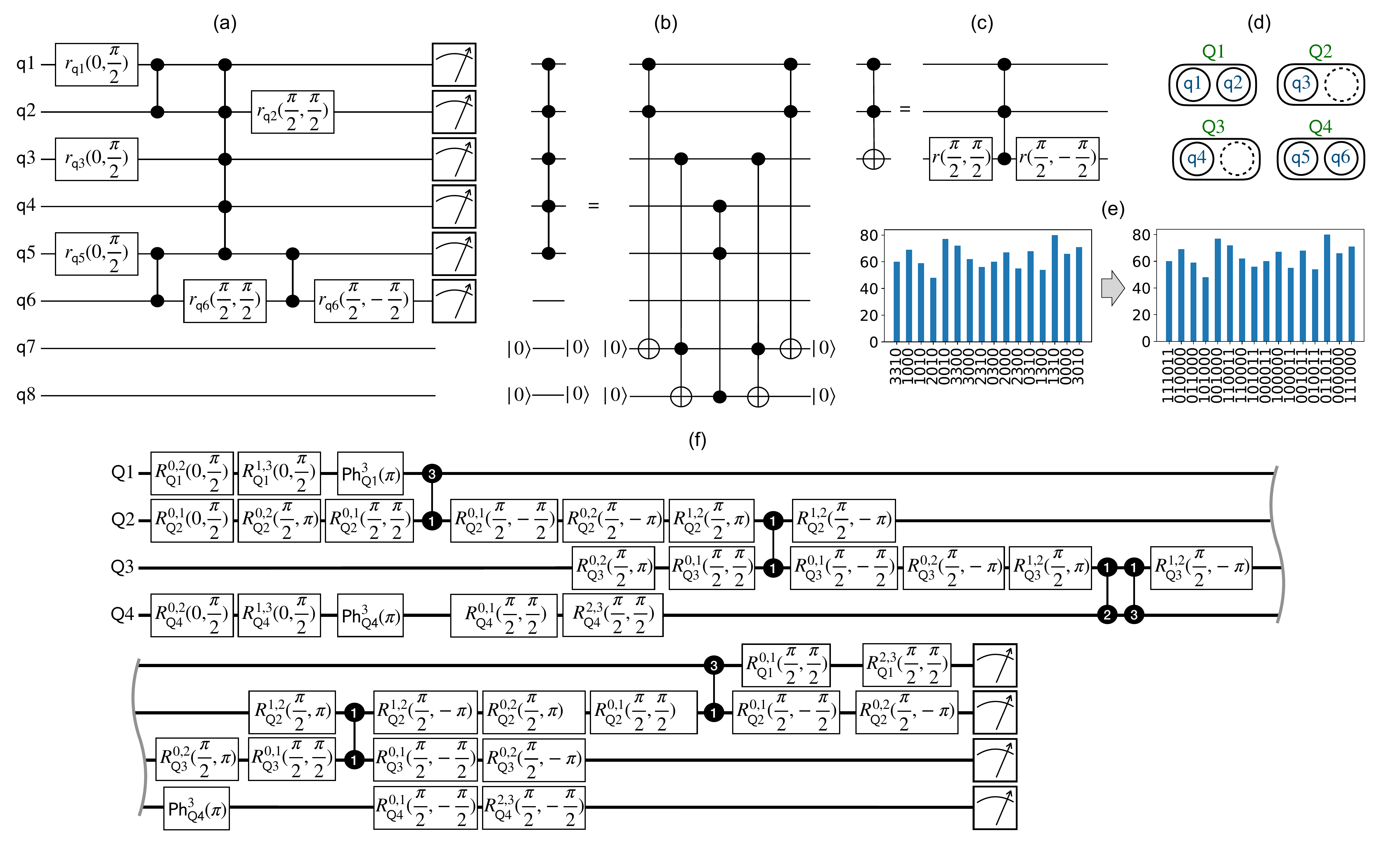}}
\vskip -3mm
\caption{
	(a) Example of a qubit circuit that is an input to the developed qudit-based transpiler.
	The circuit acts on $n=6$ qubits ${\sf q}1,\ldots,{\sf q}6$, two additional qubits ${\sf q}7, {\sf q}8$ are used for decomposing five-qubit ${\sf CZ}_{{\sf q}1,\ldots,{\sf q}5}$ gate.
	(b) Decomposition of five-qubit ${\sf CZ}_{{\sf q}1,\ldots,{\sf q}5}$ gate with two `clean' ancillas down to Toffoli gates.
	For the decomposition of Toffoli gates, the circuit identity shown in (c) and then the scheme of Fig.~\ref{fig:multi-qudit-gate}(d) can be used.
	(d) Qubit-to-qudit mapping that is used for the realization of the given qubit circuit with $m=4$ qudits of $d=4$ levels.
	(e) Equivalence of read-out results that are obtained with qudit-based emulator and post-processed outputs that can be interpreted as results of the qubit circuit implementation.
	(f) The result of qudit-based transpilation of the input qubit circuit.}
\label{fig:example}
\end{figure*}

\paragraph{Uncomputation steps (iv) and (v).}

By its construction, the implementation of Steps (i)-(iii) realizes an acquiring of the phase factor of -1 to such qudits input state, where all embedded qubits affected by the decomposed gate ${\sf CZ}_{{\sf q}1,\ldots,{\sf q}\kappa}$ are in state $\ket{1}$. 
However, we employ ancillary level $\ket{\rm a}$ of type-$\cal B$ qudits.
To remove the population from $\ket{\rm a}$ to original levels, we employ uncomputation, which is a `mirror reflection' of steps (i) and (ii).
Namely, Steps (iv) and (v) are obtained as a Hermitian conjugate of a sequence of steps (i) and (ii): their order is reversed, and each $r_{{\sf Q}j}(\varphi,\theta)$ i
s transformed to $r_{{\sf Q}j}(\varphi,-\theta)$ (note that ${\sf CZ}_{{\sf Q}j_1{\sf Q}j_2}={\sf CZ}_{{\sf Q}j_1{\sf Q}j_2}^\dagger$).
As it was already mentioned, the uncomputation also removes the relative phase between the subspace of affected qubits and the remaining subspace of type-$\cal A$ qudits, possibly acquired in Step (i).

We see that all routines during the processing of the multi-qudit case are efficient, and the resulting complexity has no more than quadratic growth with an increase the degree of the processed generalized Toffoli gate (the quadratic asymptotics can appear from the used template for ancillary-free decomposition~\cite{Barenco1995}).
That is why the whole complexity of transpiling an $n$-qubit circuit consisting of $L$ gates scales linearly $L$ and no more than polynomially with $n$.

\section{Realizing 6-qubit quantum circuit with ququarts}\label{sec:example}

As an example, we consider the realization of an $n=6$ qubit circuit, which is presented in Fig.~\ref{fig:example}(a), with a qudit-based processor consisting of $m=4$ `ququarts' (qudits with $d=4$). 
First, let us consider a straightforward implementation of the input circuit with a qubit-based processor. 
To simplify the transpiration of multi-qubit gate ${\sf CZ}_{{\sf q}1,\ldots,{\sf q}}6$, we use two additional ancillary qubits ${\sf q}7$, ${\sf q}8$.
Using schemes shown in Fig.~\ref{fig:example}(b,c) together with one from Fig.~\ref{fig:multi-qudit-gate}(d), ${\sf CZ}_{{\sf q}1,\ldots,{\sf q}6}$ can be realized with $5\times6=30$ two-qubit gates and a number of single-qubit gates.
In this way, the straightforward qubit-based decomposition of the input circuit results in $N_{\text{2-body}}^{\rm qb}=33$ two-qubit operations.
We note that no restrictions on the coupling map between qubits are considered here.

In contrast, the qubit-to-qudit mapping, shown in Fig.~\ref{fig:example}(d), allows realizing the input qubit circuit with only $N_{\text{2-body}}^{\rm qd}=6$ two-qudit gates. 
In Fig.~\ref{fig:example}(e) we show a transformation of 1024 measurement outcomes, obtained with a qudit-based classical emulator, to the read-out measurement outcomes performed in the input qubit circuit.
The transpiled qudit circuit is shown in Fig.~\ref{fig:example}(f).
One can see that the qudit-based realization provides an advantage both in the circuit width and depth.

Recall that the results of the qudit-based transpilation, shown in Fig.~\ref{fig:example}, remain also valid for initial qubit states other than $\ket{0}^{\otimes n}$.
To realize the qubit circuit with respect to another initial state using qudits, the only thing that is required is to update the initial state of qudit's register in accordance with the qubit-to-qudit mapping.
We note that within the considered mappings of the form~\eqref{eq:mapping}, any separable state of qubits corresponds to a separable state of qudits (but not vice-versa).

\section{Discussion}\label{sec:disc}

Here we stress some important points related to the developed qudit-based transpilation approach.
First, we emphasize its scalability with respect to the width and depth of a processed input qubit circuit.
The scalability is assured by the facts that (i) the complexity of transpiling of a given single-, two-, or multi-qubit gate to its qudit version grows no more than polynomially with the number of qubits affected by the gate, (ii) the complexity of transpiling the whole qubit circuit with respect to the given qubit-to-qudit mapping grows linearly with a number of gates, and (iii) a polynomial in qubit number $n$ algorithm for mapping finding, which provides an advantage (or at least doesn't make things worse) compared to a standard qubit-based transpilation, can be used.
In particular, a greedy algorithm for the mapping finder of $\mathcal{O}(n^3)$ complexity for $4\leq d\leq 7$ is shown.
Therefore, the resulting qudit-based transpilation complexity is polynomial in the number of qubits $n$ in the processed circuit and linear in the number of gates.

We also recall that in the case of $m<n\leq m\lfloor \log_2d \rfloor$, where $m$ is the number of available $d$-dimensional qudits, the developed transpilation approach makes it possible to run an $n$-qubit circuit, which cannot be launched at all with $m$ qubits.
Thus, our approach allows one to expand the range of algorithms suitable for running in terms of the required number of qubits.
On the other hand, if $m\geq n$, then the corresponding qudit circuit definitely has no more entangling gates than the transpiled in a standard qubit-based version and strictly less number of entangling gates, if there is at least one multiqubit gate in the original qubit circuit.

An important feature of the developed approach is its adaptiveness to qudit dimension in the processing of multi-qubit gates.
It allows leveraging the power of extra levels in qudits to a greater extent compared to the approaches (see, e.g.,~\cite{Mato2023graphs}), where first a given qubit circuit is transpiled down to single- and two-qubit gates, and then a qubit-to-qudit mapping is obtained.

These features together demonstrate the applicability of our approach to useful near-term quantum algorithms. 
The main application area of our approach is quantum algorithms, which typically contain multi-qubit gates. 
A clear example of two-particle gate reduction provided by qudit-based realization is Grover's search algorithm. 
As shown in Ref.~\cite{nikolaeva2023ququints},
a thousandfold reduction in entangling gate number starting from eight qubits implementation can be achieved with ququints ($d=5$). 
Multi-qubit gates are inherent in solving factorization~\cite{Shor1994} and discrete logarithm~\cite{litinski2023compute} problems.
It's worth emphasizing that decompositions of general multiqubit unitaries, e.g. Haar-random, are also based on generalized Toffoli gates~\cite{nielsen2010quantum}.
We also note that the presented qudit-based transpilation approach is promising within the employing Toffoli + Hadamard universal gate set~\cite{aharonov2003simple}, where new interesting results were reported recently~\cite{amy2023improved}. 

\section{Conclusion and outlook}\label{sec:concl}

We have presented the approach for an efficient implementation of qubit circuits with qudit-based processors. 
The proposed approach consists of finding the optimized qubit-to-qudit mapping, transpiling a qubit circuit according to this mapping, running a transpiled circuit on a qudit-based processor (or emulator), and then reassigning read-out measurement results back to the qubit-based representation.
We have developed a qudit-based transpilation algorithm with respect to a particular universal set of single-qudit and two-qudit gates and proposed an idea of a mapping finder algorithm with polynomial complexity.
Then we have shown an example of applying the
developed approach for realizing a $6$-qubit circuit with four $4$-level qudits. 
We have demonstrated that the resulting number of two-particle operations required for implementing the given circuit with qudits appears considerably smaller than the one within a straightforward qubit-based implementation.
Taking into account recent progress in improving the fidelity of qudit gates, we expect an overall increase in the resulting fidelity of implementing qubit circuits with qudits.

We note that the main goal of the current work is to provide a general approach for qubit circuit execution with qudit-based hardware.
The considered example of qudit-based transpilation has to be modified for each particular physical platform with a specific set of native gates and qudits' connection topology.
We leave these particular platform-specific problems for further consideration.
Although the realization of $d'$-ary circuits with qudits is beyond the scope of this work, their transpilation for qudit processors with $d\geq d'+1$ levels is also an improvement option for the developed qudit transpiler.

We also note that one can consider a refinement of the optimized qudit circuit criterion. 
It can be defined not only by the number of two-particle operations but also as a total qudit circuit fidelity (or its estimation), which takes into account both single-qudit and two-qudit gates fidelities.
Although recent papers demonstrate that fidelities of single- and two-qudit gates are comparable with qubit gates' fidelities, this metric allows one to more accurately take into account the effects of decoherence arising from the usage of upper levels.

\appendix

\section{Intermediate circuit in the case of incomplete set of type-${\cal A}, {\cal B}, {\cal C}$ qudits} \label{sec:special_cases}

While processing multi-qubit gates ${\sf CZ}_{{\sf q}i_1,\ldots, {\sf q}i_\kappa}$ with $|{\sf qudit{\_}set}|\geq 3$, it is possible a situation, where one or two types (${\cal A}$, ${\cal B}$, ${\cal C}$) are missing.
In this case, the described decomposition needs some slight corrections. 

If there are no type-$\cal B$ qudits, then the intermediate circuit consists of a single multi-qudit ${\sf CZ}$ gate acting on all $|{\sf qudit{\_}set}|$ qudits.
To decompose this gate we take a standard multi-qubit gate decomposition as a template as it is described in Sec. \ref{sec:type-a-cz} and \ref{sec:type-c-cz}. 
Then each qubit gate is replaced with the corresponding qudit gate(s) taking into account the type of involved qudits (type $\cal A$ or type $\cal C$).
The phase correction, discussed in Sec.~\ref{sec:type-c-cz}, has to be applied if it is necessary. 

If there are no type-$\cal A$ qudits, yet there is at least one type-$\cal B$ qudit, the ladder-like part of decomposition is started with the control on the first qudit in the state \one. 
If type-$\cal C$ qudits are also missing, then the ladder-like part of the decomposition ends with a target on the last type $\cal B$ qudit in the state \one.

\section{Alternative form of the intermediate circuit}\label{sec:alt_int_scheme}

\begin{figure}[b]
	\centering
	\includegraphics[width=\linewidth]{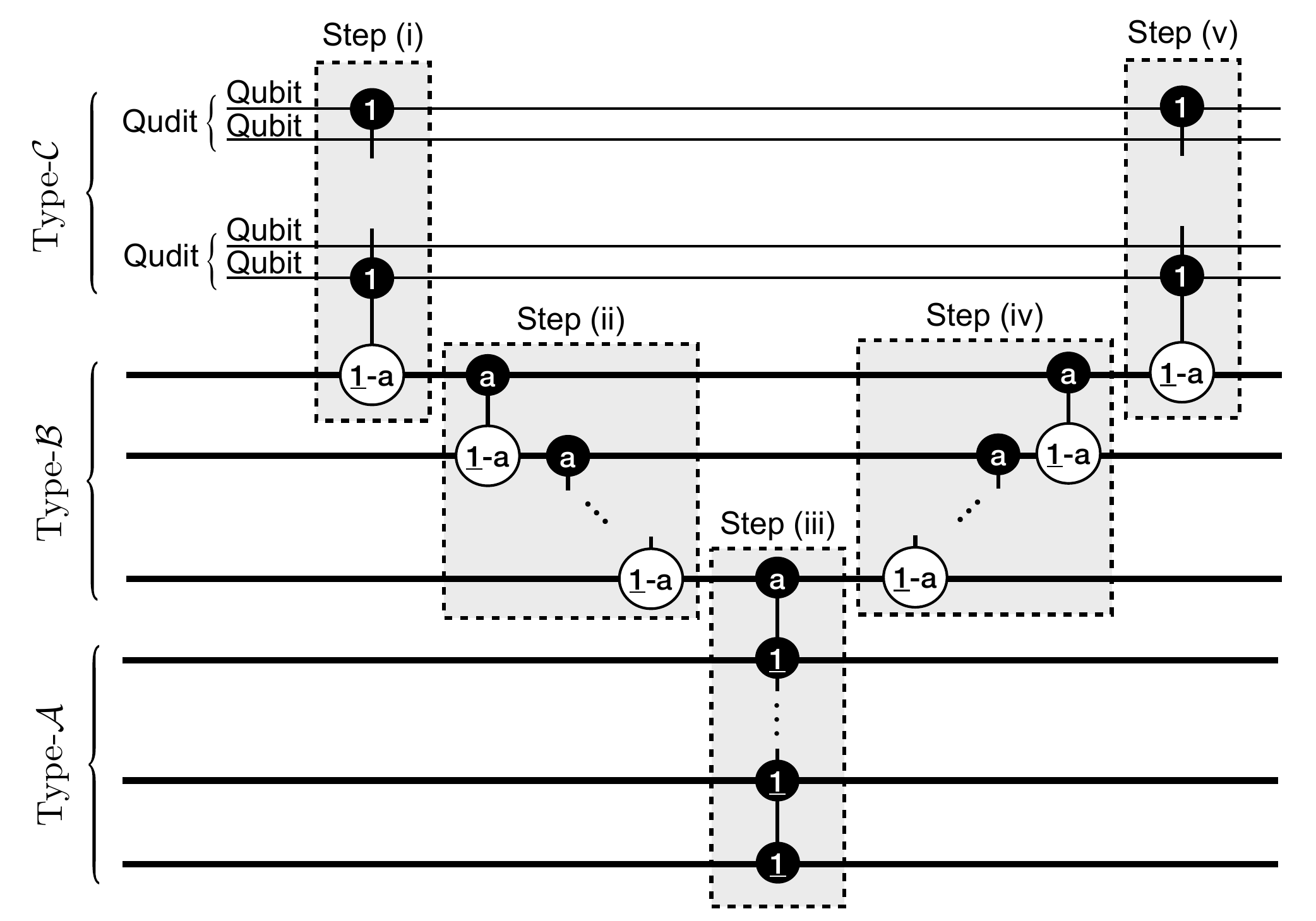}
	\caption{Alternative intermediate qudit circuit ${\sf circ}^{\text{qd-int-alt}}_{\phi}$ for the decomposition of ${\sf CZ}_{{\sf q}i_1,\ldots, {\sf q}i_\kappa}$ gate. 
	The difference from the described in the main text scheme (see Fig.~\ref{fig:multi-qudit-gate}) is the locations of swapping operations of type-$\cal A$ and type-$\cal C$ qudits.}
	\label{fig:alt_scheme}
\end{figure}

The structure of the intermediate qudit circuit ${\sf circ}^{\text{qd-int-alt}}_{\phi}$, which is presented in Fig.~\ref{fig:alt_scheme} is similar to the structure of the previously described ${\sf circ}^{\text{qd-int}}_{\phi}$.
The main difference between them is the location of type $\cal A$ and $\cal C$ qudits in the scheme. 
In the alternative version, the multi-qudit gate on type-$\cal C$ qudits is employed twice, and the multi-qudit gate on type $\cal A$ is employed once in the central part of the scheme.
We note that operations on type $\cal B$ qudits remain the same as in the previously described scheme in Fig.~\ref{fig:multi-qudit-gate}(a). 

Taking into account invariability of operations on type $\cal B$ qudits and the symmetry of $\sf CZ$ type operation in the core of type-$\cal A$ and $\cal C$ multi-qudit gates,
realization of the $\sf CZ$ type operation on  $\cal A$ and $\cal C$ qudits reduces to the procedures described in Secs.~\ref{sec:type-a-cz} and \ref{sec:type-c-cz}, correspondingly. 

\section*{Declarations}
\subsection*{Ethical Approval and Consent to participate}
Not applicable 
\subsection*{Consent for publication}
Not applicable 
\subsection*{Availability of supporting data}
Not applicable
\subsection*{Competing interests}
Owing to the employment and consulting activities of authors, A.S.N., E.O.K., and A.K.F. have financial interests in the commercial applications of quantum computing. A.S.N., E.O.K., and A.K.F. do not have any non-financial competing interests.
\subsection*{Funding}
The research is supported by the Russian Science Foundation (Grant No. 20-42-05002; qudit-based approach), the Leading Research Center on Quantum Computing (Agreement No. 014/20; implementation of quantum algorithms), and by the Priority 2030 program at the National University of Science and Technology ``MISIS'' (aspects of the transpilation scheme). 
\subsection*{Authors' contributions}
A.S.N. designed the transpilation algorithm and wrote the manuscript with support from E.O.K. and A.K.F. E.O.K. developed the theoretical formalism and general  computational framework. A.K.F. analyzed experimental limitations. All authors contributed to the final version of the manuscript. A.K.F. supervised the project.

\bibliography{bibliography-qudits.bib}

\end{document}